\documentstyle[aps,pre,epsf,floats,twocolumn,eqsecnum]{revtex}

\def \epm{{\hat\varepsilon_m}} 
\def \epinft{{\varepsilon_{T\infty}}}
\def \epinfn{{\varepsilon_{N\infty}}} 
\def \Delep{{\Delta\epsilon_{\infty}}} 
\def \Ym{{\hat Y_m}} 
\def \vm{{\hat v_m}}

\def \klp{{k_{l+}}}
\def \klm{{k_{l-}}} 
\def \ktp{{k_{t+}}} 
\def \ktm{{k_{t-}}} 
\def \krp{{k_{R+}}} 
\def \krm{{k_{R-}}}

\begin{document}

\wideabs{
\title{Critical examination of cohesive-zone models \protect\\ in the
  theory of dynamic fracture}

\author{J. S.\ Langer$^\dag$ and Alexander E.\ Lobkovsky$^\ddag$}
\address{$^\dag$ Physics Department, University of California, Santa
  Barbara, CA 93106 \\ $^\ddag$ Institute for Theoretical Physics,
  University of California, Santa Barbara, CA 93106}

\draft

\date{\today}

\maketitle

\begin{abstract}
  We have examined a class of cohesive-zone models of dynamic mode-I
  fracture, looking both at steady-state crack propagation and its
  stability against out-of-plane perturbations.  Our work is an
  extension of that of Ching, Langer, and Nakanishi (CLN), who studied
  a non-dissipative version of this model and reported strong
  instability at all non-zero crack speeds.  We have reformulated the
  CLN theory and have discovered, surprisingly, that their model is
  mathematically ill-posed.  In an attempt to correct this difficulty
  and to construct models that might exhibit realistic behavior, we
  have extended the CLN analysis to include dissipative mechanisms
  within the cohesive zone.  We have succeeded to some extent in
  finding mathematically well posed systems; and we even have found a
  class of models for which a transition from stability to instability
  may occur at a nonzero crack speed via a Hopf bifurcation at a
  finite wavelength of the applied perturbation.  However, our general
  conclusion is that these cohesive-zone models are inherently
  unsatisfactory for use in dynamical studies. They are extremely
  difficult mathematically, and they seem to be highly sensitive to
  details that ought to be physically unimportant.
\end{abstract}
}

\section{Introduction}
\label{sec:intro}

In a recent publication, Ching, Langer and Nakanishi \cite{cln}
(CLN) described a direct attempt to determine the linear stability of
mode-I fracture against bending deformations.  Their basic idea was to
use a two-dimensional cohesive-zone model of the kind introduced by
Dugdale \cite{dugdale} and Barenblatt \cite{barenblatt}, and
to compute the change in the trajectory of a crack induced by an
infinitesimally small, static, spatially oscillatory shear stress.  In
this way, they hoped to learn about the mechanisms that cause
roughening of fracture surfaces and that limit the speeds of crack
propagation \cite{fineberg,sharon}.  The major part of CLN was devoted
to the mathematically difficult job of performing the fully
elastodynamic calculation that is needed to determine this perturbed
trajectory.  CLN concluded that, for cohesive-zone models that do not
include dissipative forces, the trajectories are strongly unstable.

Our present investigation has focused, first, on a reexamination of
the mathematical problems that emerged in CLN and, second, on an
attempt to build into these models some dissipative mechanisms that
might provide more realistic descriptions of fracture stability.  Our
results have been disappointing.  The fundamental assumption in CLN
was that the Barenblatt-Dugdale cohesive-zone models, suitably
generalized to apply to bending non-steady cracks, are robust,
mathematically well-posed dynamical systems that incorporate much of
the basic physics of fracture in solids.  This assumption now seems to
be incorrect.

The cohesive-zone models that we have considered have the following
properties.

(1) The crack exists on a semi-infinite surface within an isotropic,
linearly elastic solid.  We shall discuss only two-dimensional cases
in which the crack is a line that ends at the crack tip, and the
surrounding elastic solid remains in a state of either plane stress or
plane strain.  The crack-opening displacements, that define the
fracture surfaces, are the linear elastic displacements away from the
crack line.  Note that we exclude consideration of plastic deformation
either in the bulk or on the fracture surfaces, and that we also
exclude nonlinear elasticity even near the crack tip.

(2) Cohesive forces act between the fracture surfaces and are
functions of the crack-opening displacements and their time
derivatives.  These forces vanish when the crack opening displacements
become larger than some fixed range of the interactions; thus there
should be a well defined cohesive zone near the tip of the crack
within which these forces are nonzero.  The cohesive forces may be
nonlinear functions of the displacements and may contain dissipative,
rate-dependent components; but all nonlinearities and dissipative
effects are confined within the cohesive zone.

(3) Real materials cannot support infinite stresses; therefore the
elastic stresses must be bounded at every point in the system.  As in
the original Barenblatt analysis, this finite-stress condition should
determine the structure of the cohesive zone uniquely.  It should be
analogous, for example, to the finite-stress boundary condition at the
free end of a moving string.

Implicit in the statement of this third condition is the question of
how literally to take the physics of the cohesive zone.  Our original
idea was that we should not take this aspect of the model literally at
all but, instead, should insist that the dimensions of the cohesive
zone be the smallest relevant lengths in the theory, and that its
dynamic response to external forces plays the role of phenomenological
boundary conditions to be imposed at the crack tip.  In other words,
our cohesive zone should be just a mathematical device for bridging
the gap between macroscopic elastodynamics and atomic-scale mechanisms
at the crack tip. There are other possibilities, for example, in
polymer fracture where we might develop a model of fibrils forming and
breaking within an extended process zone in front of the crack.  But,
in such circumstances, we would want to add a great deal more detailed
physics than we are prepared to consider here.

We have discovered a number of interesting but not entirely reassuring
properties of the cohesive-zone models that satisfy conditions (1)
through (3) above.  Our dissipative models, in all cases where they
turn out to be mathematically well posed, exhibit stable crack
propagation at small speeds and undergo a transition to instability at
larger speeds.  We believe that the high-speed instability is
associated with the tip-stress anomaly described by CLN.  As pointed
out in their paper, the tangential stress in all models satisfying the
above conditions exceeds the normal (opening) stress everywhere along
the crack trajectory, including at the tip, for all nonzero speeds of
crack growth.  This is a ``relativistic'' effect; that is, it is a
result of the way the stress fields transform to a frame of reference
that is moving with the crack tip at some fraction of the sound speed.
This extreme form of the conventional Yoffe \cite{yoffe} argument
seems, incorrectly as it turns out, to indicate that these models are
manifestly unstable for any choice of the cohesive forces --- that the
forward direction is never preferred by the tip stresses.  Our
introduction of a dissipative term in the cohesive shear stress does
produce stability at small speeds.  However, we have serious
reservations about this result.

One of our reservations is based on the fact that stabilization at
small crack speeds in dissipative models, that turn out to be
mathematically well-posed, depend on small-scale features within the
cohesive zone, in violation of our intuition that such features should
be irrelevant in physically sensible models.  A specific, physically
implausible aspect of our results is that increasing the strength of
the dissipative term in the cohesive shear stress seems always to
drive the response of the system in the direction of instability
rather than stability.

Our most serious concern, however, is the mathematical fragility that
we have found in these models.  We have discovered that many,
apparently credible choices of cohesive forces lead to mathematically
ill posed systems.  The equations of motion often fail to have unique,
physically acceptable solutions.  When acceptable solutions do exist,
as mentioned above, they often depend sensitively upon apparently
unphysical small-scale features within the cohesive zone.  We
conclude, therefore, that this particular class of cohesive-zone
models --- despite its popularity in recent decades --- is not
suitable for use in theories of dynamic fracture.  At the very least,
we believe that useful models will have to include plastic deformation
in extended regions outside the crack tip, and that they will have to
contain enough degrees of freedom to describe the dynamics of blunting
at the tip itself.

This report is organized as follows.  In Section \ref{sec:summary} we
summarize our reformulation of the main results of CLN.  We have tried
to make this Section readable by relegating almost all mathematical
details to a set of Appendices.  Those Appendices include analysis
which is new and possibly useful for further investigations, but which
is peripheral to the main thrust of this paper. Section
\ref{sec:steady} is a report on our efforts to include dissipative
mechanisms in the steady-state theory, an exercise that seemed to us
to be a necessary part of our investigation for reasons that we
explain there.

Our principal conclusions emerge in Section \ref{sec:first-order-sol}
where we examine the mathematical properties of the equation which
determines how an advancing crack responds to bending perturbations.
We first point out an unexpected mathematical failure of the CLN
analysis.  Next, we use a simplified mathematical version of the
problem to show how adding a dissipative term to the cohesive shear
stress may cure this problem.  We then look in detail at stability
analyses for the several dissipative fracture models whose
steady-state properties are described in Section \ref{sec:steady}.
Here we point out that one perfectly plausible model, with apparently
all of the necessary ingredients, fails (like CLN) to produce a
mathematically well defined response to bending perturbations.  We
conclude this Section, and the main body of this paper, by describing
another model that behaves in much the way we had expected from
physical considerations.  That is, it is stable at small
crack-propagation speeds, becomes unstable at larger speeds, and even,
for some values of the dissipation parameters, undergoes a Hopf
bifurcation so that the instability occurs at a well defined
wavelength.  Unfortunately, we have no reason to believe that this
model is representative of any broad class of physically motivated
fracture models, or that its behavior is a robust feature of such such
a class.

\section{Summary of previous results}
\label{sec:summary}

We start this report by summarizing the strategy and initial results
of CLN.  In a number of cases, we present those results in forms that
are different from the original versions. The reader should refer to
CLN and the Appendices in the present paper for a more complete
account of the calculations leading to these formulas.

\subsection{Geometry and boundary conditions}
\label{sec:geometry}

The first step in any stability analysis must be the choice of a
steady-state configuration whose stability is to be tested.  For
fracture problems, we need to choose a system that is translationally
invariant along the direction of crack propagation.  We then must
decide upon boundary conditions.  There are several ways to pose an
elastodynamic boundary value problem so that its solution resembles
steady-state fracture.

We might, for example, imagine that a solid containing a crack is
truly infinite in extent and is loaded in such a way as to create a
concentrated stress at the crack tip strong enough to break the
material.  If we hold the crack in place by some constraint and then
release it, the system must eventually reach steady state in a finite
region around the moving tip of the crack.  This situation is not
well-suited to a stability analysis, however, because the fracture
surfaces continue to move away from each other indefinitely far behind
the tip, and part of the elastic energy initially stored in the solid
is always being carried away to infinity.

A second mathematical device, frequently used in the literature, is to
apply moving tractions to the crack faces at some fixed distance
behind the tip so that the crack advances steadily at the desired
speed.  The steady-state solution in this case does not have a wave
that propagates to infinity; however this method of controlling the
crack extension by adjusting the loading defeats our purpose of
studying the stability of a freely advancing crack.

Therefore, like CLN, we consider a crack moving from right to left
along the centerline of an infinite elastic strip occupying the region
$(-\infty <x< +\infty, -W<y<+W)$ in the $x,y$ plane.  Far ahead of the
crack, at $x\to -\infty$, the strip is uniformly strained by an amount
$\varepsilon_{xx}=\epinft$ (the strain tangential to the crack axis),
$\varepsilon_{yy}=\epinfn$ (the normal strain) and, for the moment,
$\varepsilon_{xy}=0$.  Rigidly clamped boundary conditions at $y=\pm
W$ assure that in the steady-state all of the elastic energy
initially stored in the strip must be dissipated at the crack tip.
From the beginning of the analysis, we assume that the half-width $W$
is very much larger than any other length scale in the problem (the
wavelength of the perturbing stress or the length of the cohesive
zone), thus we carry out most of our calculations in the limit
$W\to\infty$.  However, there are several places where we need to
reintroduce the length $W$.  For example, the fully relaxed width of
the crack must scale like $W$, and the stress-intensity factor that
characterizes the forces transmitted to the crack tip is proportional
to $\sqrt{W}$.

To avoid the complication of the fixed grip boundary condition, CLN
used a mathematical technique developed by Barber, Donley and Langer
\cite{barber}.  They considered two different boundary-value
problems whose solutions are equivalent in a certain limit.  The first
is the problem described in the last paragraph in which the elastic
fields satisfy simple wave equations in the bulk and fixed-grip
boundary conditions at a distant boundary.  The second is defined in
an infinite space, but the fields satisfy a driven massive wave
equation with a small mass.  If the driving force is proportional to
the desired rigid displacement and the mass is proportional to
$1/W$, the solutions of the two boundary value problems are
equivalent in the limit of very large $W$.

\subsection{Basic structure of the stability analysis}
\label{sec:stability}

Suppose that the applied strains $\epinfn$ and $\epinft$ are such that
the unperturbed crack moves at speed $v$ in the negative $x$ direction
along the centerline of the strip. We choose our units of time so that
$v$ is measured in units of the transverse sound speed.  To examine
the stability of this crack, we compute its steady-state response to a
small (i.e.~first order) external force that produces a
spatially oscillating shear stress along the $x$-axis:
\begin{equation}
  \Sigma_{xy}^{(ext)}(x,0)= \epm\,e^{imx}.
\end{equation}
The dimensionless stress $\Sigma_{xy}$ is measured in units of twice
the shear modulus, and $\epm$ is the amplitude of the perturbation
whose wavenumber is $m$. In CLN, $\Sigma_{xy}^{(ext)}$ is defined in
the entire $x,y$ plane so that, in principle, it can be the result of
tractions applied at the edges of the strip or of material
irregularities near the centerline.  However, only the value of
$\Sigma_{xy}^{(ext)}$ on the centerline is relevant for the first
order calculation presented here.

The goal of the calculation is to compute the perturbed centerline
$y=Y_{cen}(x)$ of the resulting fracture to first order in $\epm$,
that is: 
\begin{equation}
  \label{Ycen}
  Y_{cen}(x) = \Ym\,e^{imx} \equiv
  \hat\chi_Y(m,v)\,\epm\,e^{imx}.
\end{equation}
(Throughout this paper, symbols with carets such as $\epm$ and $\Ym$
denote Fourier amplitudes.) $\hat\chi_Y$ is a complex steady-state
response coefficient that depends on the wavenumber $m$ and the
average crack propagation speed $v$. If $\hat\chi_Y$ diverges at some
$v$ and some real value of $m$, then we would conclude that the system
undergoes a change in dynamic stability at that wavenumber and speed.
More generally, poles of $\hat\chi_Y$ in the complex $m$-plane are
equivalent to stability eigenvalues.  According to (\ref{Ycen}), poles
in the lower half $m$-plane correspond to stable modes, and changes in
stability occur when poles cross the real $m$-axis.

\subsection{Crack-Opening Displacements, Stresses and
  Cohesive Forces}
\label{sec:cohesive-stresses}

Let the functions $U_N^{[\pm]}(x)$ be the normal displacements,
relative to the local orientation of the centerline $y=Y_{cen}(x)$, of
the ``upper'' $[+]$ and ``lower'' $[-]$ fracture surfaces.  Similarly,
let the $U_T^{[\pm]}(x)$ be the corresponding tangential
displacements.  Then
\begin{mathletters}
  \begin{equation}
    U_N\equiv {1\over 2}\,\left[U_N^{[+]}-U_N^{[-]}\right]
  \end{equation}
  is the crack-opening normal (mode I) displacement;
  \begin{equation}
    U_S\equiv {1\over 2}\,\left[U_T^{[+]}-U_T^{[-]}\right]
  \end{equation}
  is the crack-opening shear (mode II) displacement; and
  \begin{equation}
    \label{UT}
    U_T\equiv {1\over 2}\,\left[U_T^{[+]}+U_T^{[-]}\right]
  \end{equation}
\end{mathletters}
is the average tangential displacement.  (We shall not use $U_T$
explicitly for present purposes, but it is needed in principle for a
complete description of the crack surface.)

In a similar way, we define the normal, shear and tangential
components of the stress fields evaluated at the fracture surfaces:
\begin{mathletters}
  \begin{equation}
    \label{SN}
    \Sigma_{N}(x)\equiv
    {1\over2}\left[\Sigma_N^{[+]}(x,Y_{cen}) +
      \Sigma_{N}^{[-]}(x,Y_{cen})\right];
  \end{equation}
  \begin{equation}
    \label{SS}
    \Sigma_{S}(x)\equiv {1\over 2}
    \left[\Sigma_{S}^{[+]}(x,Y_{cen})
      +\Sigma_{S}^{[-]}(x,Y_{cen})\right];
  \end{equation}
  and
  \begin{equation}
    \label{ST}
    \Sigma_{T}(x)\equiv
    {1\over2}\left[\Sigma_{T}^{[+]}(x,Y_{cen}) +
      \Sigma_{T}^{[-]}(x,Y_{cen})\right].
  \end{equation}
\end{mathletters}
As before, the subscripts N, S, and T denote tensor components rotated
into the local orientation of the centerline $Y_{cen}$.

The cohesive stress acting between the two fracture surfaces has two
components, $\Sigma_{cN}$ and $\Sigma_{cS}$, defined in analogy to
(\ref{SN}) and (\ref{SS}) respectively.  We assume that these are
functions of the local displacements and, when we include dissipative
terms, the time derivatives of these displacements:
\begin{mathletters}
  \begin{equation}
    \label{ScN}
    \Sigma_{cN}(x)=\Sigma_{cN}[U_N(x),\dot U_N(x)];
  \end{equation}
  \begin{equation}
    \label{ScS}
    \Sigma_{cS}(x)=\Sigma_{cS}[U_N(x),\dot 
    U_N(x),U_S(x),\dot
    U_S(x)].
  \end{equation}
\end{mathletters}
In principle, we could include higher time derivatives in these
equations. Moreover, in an even more general theory, this local
definition of the cohesive forces may need to be replaced by a
nonlocal form in which these forces depend on the crack opening
displacements everywhere in the plastic zone. We see no compelling
physical reason for either of these possible extensions of the present
formulation.

In our linear response theory, the shear displacements $U_S$ must be
infinitesimally small quantities which vanish when the crack is moving
steadily along a straight line.  For reasons of symmetry, these
quantities can enter the normal cohesive stress only at second order
and must therefore be neglected.  Similarly, the cohesive shear stress
must be linear in $U_S$ and its time derivatives; and it should not
depend directly on $U_S$ but, rather, on the shear angle $\Theta$
defined by 
\begin{equation}
  \label{Thetadef}
  {U_S(x)\over U_N(x)}\equiv\tan \Theta(x)\cong \Theta(x).
\end{equation}
Without loss of generality, therefore, we can write (\ref{ScS}) in the
form
\begin{equation}
  \label{ScSeta}
  \Sigma_{cS}= \Sigma_{cN}[U_N(x),\dot
  U_N(x)]\,\Bigl(\Theta(x)+\eta \dot\Theta(x)\Bigr),
\end{equation}
where $\eta$ is a dissipation coefficient and $\dot\Theta$ denotes the
time derivative of $\Theta$.  Our choice of unity for the coefficient
of the first factor $\Theta$ in (\ref{ScSeta}) corresponds to the CLN
central-force assumption.  That is, apart from the rate-dependent
factor $\eta \dot\Theta$, the cohesive stress behaves like a central
force acting between opposite points on the fracture surfaces.

A general feature of the cohesive stress in these models is that both
its normal and shear components must vanish when the crack is fully
open, that is, when $U_N(x)$ exceeds the range of the cohesive
interactions.  We define the cohesive zone to be that region near the
tip of the crack, $0 \le x \le \ell$, within which the cohesive forces
are nonzero.  The length of the cohesive zone, $\ell$, is a dynamic
quantity that depends on the state of motion of the crack.  It appears
frequently throughout this analysis.

\subsection{Steps in the CLN Analysis}
\label{sec:steps}

The CLN analysis starts by transforming the equations of linear
elasticity into a frame of reference moving in the negative
$x$-direction at a speed such that the tip of the crack is always at
$x'=0$, that is, $x=x'+x_{tip}(t)$, where
\begin{equation}
  \dot x_{tip}(t)= - v - \vm e^{-imvt}.
\end{equation}
This transformation into a non-uniformly moving frame is essential
because it allows us to deal non-perturbatively with the various
mathematical singularities that occur at the crack tip. In principle,
the new frame should be chosen so that the tip is also at $y'=0$ in
the moving $x',y'$ plane. To first order in the perturbation $\epm$,
however, this extra part of the transformation turns out to be
unnecessary. To simplify notation throughout the rest of this
analysis, we simply drop the apostrophes after performing the
transformation.

The next step in the CLN analysis is to write down formal solutions of
the equations of linear elasticity separately in the two regions of
the $x,y$ plane above and below $Y_{cen}(x)$, and then to evaluate the
unknown coefficients that occur in those solutions by imposing
boundary conditions on the centerline.  Ahead of the crack tip, the
centerline is purely fictitious and the boundary conditions are simply
statements that the stresses and displacements must be continuous
there.  Behind the tip, on the other hand, these boundary conditions
are statements about tractions on the fracture surfaces.  That is, the
stresses at the fracture surface, $\Sigma_N(x)$ and $\Sigma_S(x)$,
must be balanced by the cohesive stresses in the cohesive zone and
must vanish where the crack is open behind that zone.

It is also necessary, at this stage of the analysis, to separate the
problem into parts which are zeroth order and first order in the
perturbation $\epm$. CLN introduced an extra subscript, $0$ or $1$, to
indicate the order at which each function was being computed.  That
will not be necessary here.  All symbols with subscripts $N$ are
zeroth order in $\epm$, and all with subscripts $S$ are first order.

The combination of elasticity and boundary conditions produces a set
of Wiener-Hopf equations that can be solved for the unknown stresses
ahead of the tip and the unknown crack-opening displacements behind
it. At zeroth order in $\epm$, the equation that relates $U_N(x)$ and
$\Sigma_N(x)$ is a restatement of the conventional, steady-state,
cohesive-zone model. As in the original analysis of Barenblatt, the
condition that the normal stress $\Sigma_N(x)$ remains finite at the
crack tip is sufficient to produce a unique solution.  At first order
in $\epm$, the equation involving $U_S(x)$ and $\Sigma_S(x)$ contains
the information that is needed to find $\hat\chi_Y(m,v)$.  Again, the
condition that $\Sigma_S(0)$ remains finite appears to produce a
mathematically well defined equation for $U_S$.  As we shall see, the
situation is not so simple.

\subsection{Zeroth Order Problem}
\label{sec:zero}

The zeroth order, steady-state problem is most conveniently cast in
the form of a nonlinear, singular integral equation for the
crack-opening displacement $U_N(x)$.  (For a more detailed derivation,
see Appendix A.)  This equation is:
\begin{equation}
  \label{U'x}
  {dU_N\over dx}={1\over\pi b(v)}\int_0^{\ell}dy\,\sqrt{x\over y}\, 
  {1 \over x-y}\,\Sigma_{cN}[U_N(y)];
\end{equation}
or, equivalently,
\begin{eqnarray}
  \label{UNx}
  && U_{N}(x) \equiv {1\over\pi b(v)}\, \int_0^{\ell} dy\, Z(x,y)\,
  \Sigma_{cN}[U_N(y)] \cr
  && = {1\over\pi b(v)} \int_0^{\ell}dy \left[2\sqrt{x\over 
      y}-\ln\left|{\sqrt y +\sqrt x\over\sqrt y - \sqrt x}\right|
  \right] \Sigma_{cN}[U_N(y)],
\end{eqnarray}
where the first version of the right-hand side serves as a definition
of the kernel $Z(x,y).$  Eqs.~(\ref{U'x}) and (\ref{UNx}) are valid
everywhere in the region $0\le x \ll W$.

Throughout this paper, all singular integrals such as that appearing
in (\ref{U'x}) denote Cauchy principal values. For simplicity, we have
suppressed the argument $\dot U_N = v\,dU/dx$ in writing the function
$\Sigma_{cN}$.  The quantity
\begin{equation}
  b(v)={2\over\beta_lv^2}(\beta_l\beta_t-\beta_0^4);
\end{equation}
plays an important role in much of what follows. Here,
\begin{equation}
  \beta_l^2\equiv 1 - {v^2\over\kappa};\qquad 
  \beta_t^2\equiv
  1-v^2;\qquad \beta_0^2\equiv 1-{v^2\over 2};
\end{equation}
and the parameter $\kappa$ is the square of the ratio of the
longitudinal to transverse sound speeds. $b(v)$ vanishes at the
Rayleigh speed $v_R$, $b(v_R)=0$.

The upper limit of integration in (\ref{U'x}) is the length of the
cohesive zone $\ell$, defined in the paragraph following
Eq.~(\ref{ScSeta}).  Let $\delta$ be the maximum range of the cohesive
force.  Then $\ell$ is determined by the relation
\begin{equation}
  \label{elldelta}
  U_N(\ell)=\delta.
\end{equation}

Equation (\ref{U'x}) must be supplemented by the Barenblatt condition
(non-divergent stress at $x=0$): 
\begin{equation}
  \label{BCond}
  \Sigma_{N\infty} = \left[\kappa\over 2\pi\, b(v)\, W\right]^{1/2}
  \int_0^{\ell}{dx\over\sqrt{x}}\, \Sigma_{cN}[U_{N}(x)]
\end{equation}
where,
\begin{equation}
  \Sigma_{N\infty}\equiv
  \left(\kappa\over2\right)\epinfn+\left({\kappa\over2}-
    1\right)\epinft
\end{equation}
is the normal stress infinitely far ahead of the crack determined by
the applied strains $\epinfn$ and $\epinft$.

We also need to know that the energy dissipated per unit length of
crack extension is 
\begin{equation}
  \label{Gammadef}
  \Gamma_{frac}\equiv 
  \int_0^{\ell}dx\,\Sigma_{cN}[U_N(x)]\,{dU_N\over dx}.
\end{equation}
Because the linearly elastic material outside the cohesive zone is
energy-conserving, the flux $\Gamma_{frac}$ defined in this way
accounts for the entire conversion of elastic energy to fracture
energy in this system.  If $\Sigma_{cN}$ includes dissipative effects,
$\Gamma_{frac}$ will be a function of the crack speed $v$.  If we use
first (\ref{U'x}) and then (\ref{BCond}) in (\ref{Gammadef}), we find
\begin{eqnarray}
  \label{Gammaexp}
  \nonumber
  \Gamma_{frac}(v) & = & {1 \over \pi b(v)} \int_0^{\ell} dx 
  \int_0^{\ell}
  dy\, \Sigma_{cN}(x) \sqrt{x\over y}\, {1\over x-
    y}\,\Sigma_{cN}(y)\\
  & = &{1\over\pi b(v)} \left(\int_0^{\ell}{dx \over \sqrt 
      x}\,
    \Sigma_{cN}(x) \right)^2 = {W \over 
    \kappa}\,\Sigma_{N\infty}^2.
\end{eqnarray}
The last equality restates the fact that all the elastic energy
initially stored in the strip is dissipated at the crack tip.  In the
limit $v\to 0$, or in the case where the cohesive stress is purely
non-dissipative, $\Sigma_{N\infty}$ must be equal to the Griffith
threshold stress $\Sigma_G$: 
\begin{equation}
  \label{SigmaG}
  \Sigma_{N\infty} = \Sigma_{G} = \left(\kappa\, \Gamma_0 
    \over W
  \right)^{1/2},
\end{equation}
where $\Gamma_0$ denotes the non-dissipative (threshold) fracture
energy.

It will be important to recognize that, outside the cohesive zone
where $\ell \ll x \ll W$, (\ref{UNx}) shows the conventional $\sqrt x$
behavior of the crack opening displacement.  Only the first term in
$Z(x,y)$ contributes in this limit. Thus, with (\ref{BCond}):
\begin{equation}
  \label{Usqrtx}
  U_N(x) \approx {2\sqrt{x}\over \pi\,b(v)}\int_0^{\ell}
  {dy\over\sqrt{y}} \, \Sigma_{cN}(y) = \left[{8W \over \pi
      \kappa b(v)}\right]^{1/2}\!\! \Sigma_{N\infty} \sqrt{x}.
\end{equation}
Note that all of these results so far are completely independent of
our specific choice of cohesive stresses.

We shall find it convenient to use a scaled version of the above
equations. Indeed, these scaling transformations are needed in order
to uncover the mathematical structure of the problem. Let $\Sigma_0$
be some characteristic stress, perhaps the yield stress at the crack
tip.  Then define 
\begin{equation}
  \label{zetadef}
  \xi = {x \over \ell},\quad \zeta(\xi)={\pi b(v)\over 
    \ell\,\Sigma_0}\,U_N(x);
\end{equation}
and
\begin{equation}
  \label{sigmadef}
  \Sigma_{cN}[U_N]\equiv\Sigma_0\,\tilde\sigma_{cN}[U_N(x)] \equiv
  \Sigma_0\,\sigma_{cN}[\zeta(\xi)].
\end{equation}
Again, we suppress the explicit dependence of $\sigma_{cN}$ on the
derivative of $\zeta$, but must remember in later applications that it
may be present.  With this notation, (\ref{U'x}) becomes
\begin{equation}
  \label{zeta'xi}
  {d\zeta\over d\xi}=\int_0^1 d\xi'\,\sqrt{\xi\over \xi'}\, 
  {1\over \xi-\xi'}\,\sigma_{cN}[\zeta(\xi')],
\end{equation}
and the Barenblatt condition (\ref{BCond}) is
\begin{equation}
  \Sigma_{N\infty} = \left[\kappa \ell \over 2\pi\, b(v)\,
    W\right]^{1/2}\, \Sigma_0\,\int_0^1{d\xi\over\sqrt{\xi}}\,
  \sigma_{cN}[\zeta(\xi)]
\end{equation}
The relation (\ref{elldelta}) that determines $\ell$, combined with
(\ref{zetadef}), becomes 
\begin{equation}
  \label{ellzeta}
  \ell = \delta {\pi b(v)\over \Sigma_0\,\zeta(1)}.
\end{equation}
It is natural, therefore, to measure $\ell$ and other lengths in the
units of $\delta$.  We return to this steady-state problem in Section
\ref{sec:steady}.

\subsection{First-Order Problem}
\label{sec:first}

The CLN result for the response coefficient $\hat \chi_Y(m,v)$ has the
form 
\begin{equation}
  \label{chi1}
  -{\epm \over \Ym} = -\hat \chi_Y^{-1}(m,v) = im \Delep + \tilde{\cal
    D}_0(m,v)+ \tilde{\cal D}_1(m,v)
\end{equation}
where
\begin{equation}
  \Delep = \epinfn - \epinft. 
\end{equation}
It is useful for mathematical purposes --- almost essential, in fact
--- to restrict our attention to the limiting case $m\ell\ll 1$.  That
is, we look only at perturbations whose wavelengths are very much
larger than $\ell$, the length of the cohesive zone.  As discussed in
Section \ref{sec:intro}, we expect $\ell$ to be smaller than any other
relevant length in the system, including $2\pi/m$. Thus this
assumption seems reasonable for both mathematical and physical
reasons.  We shall see, however, that it limits our ability to explore
the analytic properties of $\hat \chi_Y(m,v)$.

In the small-$m$ limit, the function $\tilde{\cal D}_0(m,v)$ can be
written in the form [see Eq.~(\ref{D0formapp})]
\begin{equation}
  \label{D0form}
  \tilde{\cal D}_0(m,v) \cong im (-imW)^{1/2} 
  \Sigma_{N\infty} \, {\cal
    D}_0(v).
\end{equation}
The quantity ${\cal D}_0(v)$ is a function only of $v^2,$ even for
cases in which the cohesive stress contains dissipative terms.  ${\cal
  D}_0(v)$ has the limiting value
\begin{equation}
  \label{D00}
  \lim_{v \to 0}{\cal D}_0(v) = \left(\kappa - 1 \over 2 
    \kappa^2
  \right)^{1/2}.
\end{equation}
It is a positive, monotonically increasing function in $0\le v < v_R$,
and it diverges as $v$ approaches the Rayleigh speed $v_R$.

The first two terms on the right-hand side of (\ref{chi1}) make no
reference to the cohesive shear stresses defined in (\ref{ScS}) or
(\ref{ScSeta}).  CLN pointed out that these first terms can be
obtained by neglecting the cohesive shear stresses but constraining
the elastic stress $\Sigma_S$ to vanish at the crack tip --- in effect
imposing a ``$K_{II}=0$'' constraint.  This first part of the response
function therefore constitutes a generalization to non-zero velocities
of the Cotterell and Rice (CR) \cite{CR} quasi-static theory of
crack extension.  The quantity $-\Delep$ is precisely the CR ``$T$
stress''; and the factor $(-imW)^{1/2}$ in (\ref{D0form}), when
included in (\ref{chi1}), produces a weak ($W$-dependent) instability
for negative values of $\Delep$.

All of the effects of the cohesive shear stress are contained in the
third term on the right-hand side of (\ref{chi1}).  In analogy to
(\ref{D0form}), and again in the small-$m$ limit, we can write this
quantity in the form [see Eq.~(\ref{D1approx})]
\begin{equation}
  \label{D1form}
  \tilde{\cal D}_1(m,v) \cong -m^2 \ell (-imW)^{1/2}
  \Sigma_{N\infty} {\cal D}_1(v) \! \int_0^1 \! {d\xi \over
    \sqrt{\xi}} {\cal S}_{cS}[a(\xi)].
\end{equation}
Here, ${\cal D}_1(v)$, like ${\cal D}_0(v)$, is a function only of
$v^2$ and has the same limiting value (\ref{D00}) at $v=0$.  It too is
a monotonically increasing function of $v$ which diverges, more
strongly than ${\cal D}_0(v)$, as $v\to v_R$.

The dimensionless quantity ${\cal S}_{cS}$ is proportional to the
cohesive shear stress given in (\ref{ScSeta}).  It is defined by
\begin{equation}
  \label{calS}
  {\cal S}_{cS}[a(\xi)]=\sigma_{cN}[\zeta(\xi)]\,\tau_{cS}[a(\xi)]
\end{equation}
where
\begin{equation}
  \label{taudef}
  \tau_{cS}[a(\xi)] = \left(1+im\ell{\eta v \over \ell}\right)\,a(\xi)
    + {\eta v\over \ell}\, {da \over d\xi}.
\end{equation}
Here, we have transformed the linear form $\Theta+\eta\dot\Theta$ into
the moving frame and have replaced the shear angle $\Theta$ by a new
function denoted $a(\xi)$.  The latter quantity contains much of the
interesting structure of the problem.  Apart from multiplicative
constants [shown in Eq.~(\ref{Afactors})], $a(\xi)$ is the ratio of
the shear angle $\Theta$ to the displacement of the perturbed
centerline $Y_{cen}$; that is, it tells us how much shear is induced
by the bending of the crack within the cohesive zone.  Note that we
have kept a term $im\eta v$ in (\ref{taudef}); the dissipation length
$\eta v$ is not necessarily small compared to $\ell$, and thus may be
comparable to $2\pi/m$.

The equation that determines $a(\xi)$ is deduced from the Wiener-Hopf
equation that relates $U_S$ to $\Sigma_S$, supplemented by the
condition that $\Sigma_S(x)$ be non-divergent at $x=0$.  [See the
analysis leading to Eq.~(\ref{Aeqn}).] The result is a linear,
inhomogeneous singular integral equation: 
\begin{eqnarray}
  \label{aeqn}
  {3 \over 2} \sqrt{\xi} &=& {d\over d\xi} [\zeta(\xi) a(\xi)] -\cr
  {\beta_t \over \beta_l} \int_0^1 d\xi' {d \over d\xi'} &&
  \!\!\!\!\!\!\!\!\!  \left[\ln\left|{\sqrt{\xi} + \sqrt{\xi'} \over
        \sqrt{\xi} - \sqrt{\xi'}}\right| e^{-im\ell(\xi -
      \xi')}\right] {\cal S}_{cS}[a(\xi')].
\end{eqnarray}
This equation, in the non-dissipative case, is equivalent to
Eq.~(5.23) in CLN. We discuss it in detail in Section
\ref{sec:first-order-sol} of the present paper.

To conclude this summary, we rewrite (\ref{chi1}) in the form
\begin{equation}
  \label{chi-scaled}
  -\hat\chi_Y^{-1}(m,v) = im\Delep + im(-imW)^{1/2} 
  \Sigma_{N\infty} {\cal D}_0(v) \Phi(m,v),
\end{equation}
where
\begin{equation}
  \label{Phidef}
  \Phi(m,v) = 1 + im \ell \, {{\cal D}_1(v) \over {\cal 
      D}_0(v)}\, \int_0^1 {d\xi \over \sqrt{\xi}} {\cal
    S}_{cS}[a(\xi)].
\end{equation}
Note that $\hat\chi_Y^{-1}(m,v)$ is proportional to $im$.  This is a
consequence of translational symmetry in the $y$-direction; it is the
slope of the perturbed trajectory, that is, $im\,\Ym$, and not its
magnitude that responds to the perturbation $\epm$.  More importantly,
the second term on the right-hand side of (\ref{Phidef}), the term
that derives from the cohesive shear stress, is explicitly
proportional to $im\ell$.  We trace this factor to the symmetry
requirement that the inclusion of the cohesive shear stress, that
is an odd function of the shear angle $\Theta$, can only affect the
curvature of $Y_{cen}$ and not its slope.

At first glance, it would seem that we should drop the $m\ell$ term in
(\ref{Phidef}) to be consistent with our small-$m$ approximation.  In
that case, we would have just $\Phi = 1$ and would again recover the
Cotterell and Rice theory.  The main result of CLN, however, was based
on the observation that $a(\xi)$ may diverge at small values of
$m\ell$ and $v$, and that the resulting behavior of $\Phi(m,v)$ in the
complex $m$-plane may completely change the conclusions of CR.  For
similar reasons, we have retained the factor $e^{-im\ell(\xi-\xi')}$
in the integral kernel in Eq.~(\ref{aeqn}), because this factor can
play a role at first order in $m\ell$.  Other small-$m$ corrections to
the kernel are of order $mv\ell$, and thus are doubly small near
$m=v=0$.

Finally, we remark that the ratio ${\cal D}_1(v)/{\cal D}_0(v)$ in
(\ref{Phidef}) is unity at $v=0$ and diverges as $v\to v_R$.  The
functions ${\cal D}_1(v)$ and ${\cal D}_0(v)$ are defined in detail in
Eqs.(\ref{D0exp}) and (\ref{D1exp}).  For present purposes, we need to
note only that this factor enhances the effect of the cohesive shear
stresses at large velocities.

\section{Steady-state, unperturbed solutions}
\label{sec:steady}

CLN based their calculations entirely on the standard Dugdale model
for which
\begin{equation}
  \label{DD}
  \Sigma_{cN}[U_N(x)] = \cases{
    \Sigma_0 & for $0 < U_N(x) < \delta$,\cr
    0 & for $U_N(x) > \delta$.\cr}
\end{equation}
Here, $\Sigma_0$ is the yield stress that we introduced in
(\ref{sigmadef}), and $\delta$ is the range of the cohesive force
defined in (\ref{elldelta}). Although Dugdale interpreted $\Sigma_0$
as the plastic flow stress, there is in fact no rate dependent
dissipation in this model; all of the stored elastic energy must be
converted into the fracture energy $\Gamma_0=\delta \Sigma_0$.  As a
result, the crack can advance at any velocity in the range $0\le v <
v_R$ at precisely the Griffith threshold given in (\ref{SigmaG}). At
higher driving forces, only propagation at the Rayleigh speed $v_R$
can occur with a fraction of the stored elastic energy radiated to
infinity.

As pointed out by CLN, the lack of dissipation in the steady-state
model is problematic.  A linear stability theory is an analysis of the
first-order response of a system to some change in the driving force,
and such a calculation may not make sense if only one driving force is
allowed.  The CLN procedure seemed permissible because the only
perturbations being considered were shear stresses that bent the crack
but did not accelerate it.  Once we discovered other difficulties in
CLN, however, we realized that we would have to make sure that those
problems were not related to this unrealistic aspect of the
steady-state behavior.  Accordingly, we have examined several
dissipative steady-state models, and summarize our results in the next
few paragraphs.  In short, we find that adding dissipation to the
steady state makes some interesting differences in the stability
theory but does not make qualitative changes in its mathematical
structure.

To start, we need to translate some formulas pertaining to the Dugdale
model into the scaled variables introduced in Eqs.~(\ref{zetadef}) and
(\ref{sigmadef}).  Because the function $\sigma_{cN}[\zeta(\xi)]$ is
unity throughout the cohesive zone $0<\xi<1$, we can immediately
integrate (\ref{zeta'xi}) twice to find 
\begin{equation}
  \label{zetaDD}
  \zeta(\xi) = 2\sqrt{\xi}
  -(1-\xi)\ln\left({1+\sqrt{\xi}\over 1-\sqrt{\xi}}\right).
\end{equation}
We then compute the length of the cohesive zone $\ell$ from
(\ref{ellzeta}), using $\zeta(1)=2$:
\begin{equation}
  \label{ellDD}
  \ell={\pi b(v)\delta\over 2\Sigma_0},
\end{equation}
which tells us that $\ell$ is a $v$-dependent quantity that is of
order $\delta/\Sigma_0$ at small speeds and vanishes at the Rayleigh
speed.

The conceptually simplest way of adding dissipation to the Dugdale
model was suggested by Glennie \cite{glennie}, who modified
(\ref{DD}) by writing:
\begin{mathletters}
  \begin{equation}
    \label{DDlambda}
    \Sigma_{cN}[U_N(x),\dot U_N(x)]=\cases{
      \Sigma_0\,(1 + \lambda\dot U_N), & $0 < U_N(x) < \delta$ \cr
      0, & $U_N(x) > \delta$; \cr}
  \end{equation}
  or, in terms of the scaled variables,
  \begin{equation}
    \label{DDlambdascaled}
    \sigma_{cN}[\zeta(\xi)]=\cases{
      1 + \tilde\lambda\,\displaystyle{d\zeta\over d\xi},
      & $0 < \zeta(\xi) < \tilde\delta$ \cr 
      0, & $\zeta(\xi) > \tilde\delta.$\cr}
  \end{equation}
\end{mathletters}
Here,
\begin{equation}
  \label{tildelambda}
  \tilde\lambda={\lambda v \Sigma_0\over\pi b(v)};\quad 
  \tilde\delta= {\pi b(v)\delta\over \Sigma_0\ell}.
\end{equation}
Using (\ref{DDlambdascaled}) in (\ref{zeta'xi}), we obtain an equation
that can be solved analytically by standard techniques
\cite{muskhelishvili}.  The conditions that $\zeta(\xi)$ must vanish
at $\xi = 0$ and be positive for $\xi > 0$ are sufficient to determine
the solution uniquely.  A compact form of the result has been given by
Freund \cite{freund}: 
\begin{equation}
  \label{zetaGlennie}
  {d\zeta \over d\xi} = \cos\vartheta\,{\xi^{\vartheta/\pi+1/2} \over
    (1-\xi)^{\vartheta/\pi}} \int_0^1{ds \over \sqrt{s}}\,
  {(1-s)^{\vartheta/\pi} \over 1 - s\xi},
\end{equation}
where
\begin{equation}
  \tan\vartheta = \pi\tilde\lambda={\lambda v \Sigma_0 \over b(v)}.
\end{equation}

For the most part, the properties of this solution are what one might
expect for a model of this kind.  Near the Griffith threshold,
(\ref{SigmaG}), the crack speed $v$ rises linearly from zero as
a function of the stress above the threshold with a slope inversely
proportional to the dissipation coefficient $\lambda$.  For large
values of $\tilde\lambda$, (that is, for large $\lambda v$ or for $v$
close to $v_R$ where $b(v)$ becomes small), the length of the cohesive
zone $\ell$ goes to a finite value and the energy flux becomes
\begin{equation}
  \Gamma_{fract}(v) \approx {\pi^2\over 2}\, \Gamma_0\, \tilde
  \lambda(v) = {\pi\lambda\Sigma_0\over 2}\,\Gamma_0\, {v \over
    b(v)}.
\end{equation}
This formula illustrates the behavior of such models in the limit of
vanishing dissipation.  When $\lambda$ is very small, $v$ must jump
quickly from $0$ to values near $v_R$ for a very small increment in
$\Gamma_{fract}$ above the Griffith threshold.

There is, however, one fatal deficiency of this model. For any nonzero
value of $\lambda v$, the stress diverges at $\xi=1$; and we cannot
allow divergent stresses in cohesive-zone models.  The divergence is
clear in the figures published by Glennie and Freund.  To see it from
(\ref{zetaGlennie}), note that the integral over $s$ is well defined
at $\xi=1$ for $\gamma>0$ but the prefactor diverges there.  Thus
$d\zeta/d\xi$ and, accordingly, $\sigma_{cN}[\zeta(\xi)]$ are
unbounded at the back end of the cohesive zone.  We therefore must
modify this form of the cohesive stress before using it in a stability
calculation.

We have examined two qualitatively different forms of the cohesive
stress, both of which cure the unphysical singularity at $\xi=1$.  The
first was motivated by the idea that the linear, velocity-dependent
stress shown in (\ref{DDlambdascaled}) probably ought to saturate at
arbitrarily high opening rates. Thus we write
\begin{equation}
  \label{rate-saturate}
  \sigma_{cN}[\zeta(\xi)] = {1 + \tilde\lambda_1\, \displaystyle
    {d\zeta \over d\xi} \over 1 + \tilde \lambda_2\, \displaystyle
    {d\zeta \over d\xi}}
\end{equation}
for $0 < \zeta(\xi) < \tilde\delta$.  The parameters $\tilde\lambda_1$
and $\tilde\lambda_2$ are defined in direct analogy with
(\ref{tildelambda}), with the ``bare'' parameters now named
$\lambda_1$ and $\lambda_2$. The ratio $\lambda_1/\lambda_2$
determines the highest possible dissipative stress whereas the
difference $\lambda_1 - \lambda_2$ is the effective dissipation
coefficient for small opening rates.

Another feature of the Dugdale and Glennie cohesive-zone models that
may be unphysical is the discontinuity in the stress at the back end
of the zone, $\xi=1$.  A simple way to achieve continuity in the
cohesive stress without introducing any new parameters is to replace
Dugdale's square-law stress function (\ref{DD}) by a triangular
function that vanishes at $\xi=1$.  That is,
\begin{equation}
  \label{continuous}
  \sigma_{cN}[\zeta(\xi)]=\left[1 - {\zeta(\xi)\over\tilde\delta}
  \right]\, \left(1 +\tilde\lambda\,{d\zeta\over d\xi}\right)
\end{equation}
for $0 < \zeta(\xi) < \tilde\delta$.

Neither of the latter two forms of the cohesive stress produces an
analytically solvable version of (\ref{zeta'xi}), thus we have
resorted to numerical methods.  In the process, we have discovered
that numerical solution of singular integral equations is a difficult
and subtle enterprise.  Appendix C contains a summary
of our strategy and some remarks about the mathematical nature of
these problems.

\begin{figure}[htbp]
  \centerline{\epsfxsize=3.3in \epsfbox{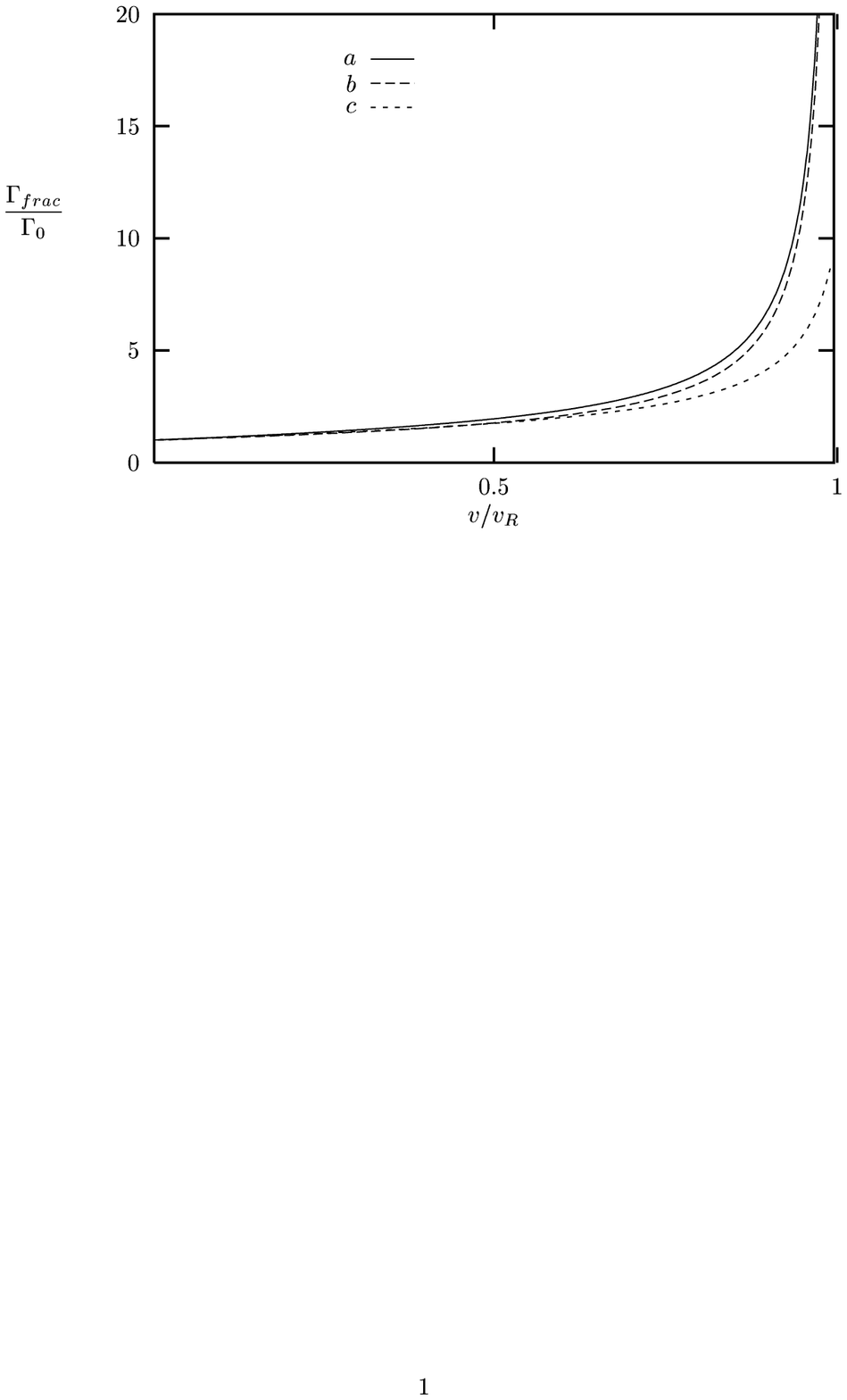}}
  \vspace{0.1in}
  FIG.~1. The theoretical fracture toughness computed for 
  three different normal cohesive stresses.  The Glennie cohesive
  stress (curve $a$) with $\lambda\Sigma_0=1$ produces results
  similar to the model with the continuous cohesive stress and the
  same value of $\lambda$ (curve $b$).  The model with the
  saturating cohesive stress (curve $c$) with
  $\lambda_1\Sigma_0=1$ and $\lambda_2\Sigma_0=0.1$ leads to a
  finite fracture toughness at $v=v_R.$ 
  \label{fig:gammas}
\end{figure}

Our numerical analysis indicates that both of the new models, defined
by Eqs.~(\ref{rate-saturate}) and (\ref{continuous}), have
non-divergent, physically acceptable solutions.  In the saturating
case, (\ref{rate-saturate}), the length of the cohesive zone $\ell$
vanishes and $\Gamma_{fract}(v)$ reaches a finite limit at $v_R$. This
means that steady-state fracture is possible only for a finite range
of driving forces in excess of the Griffith threshold.  For the
continuous cohesive stress shown in (\ref{continuous}), on the other
hand, the length of the cohesive zone $\ell$ increases to some
limiting value and $\Gamma_{fract}(v)$ diverges like $(v_R-v)^{-1}$
near $v_R$ as in the Glennie model with the discontinuous cohesive
stress.  We show some representative graphs of $\Gamma_{fract}(v)$ in
Figure \ref{fig:gammas}.

\section{First-order solutions}
\label{sec:first-order-sol}

We turn now to the main effort of this investigation, an analysis of
the integral equation (\ref{aeqn}) which should determine the linear
response of the crack to bending perturbations.  To start, we expand
the kernel in (\ref{aeqn}) to first order in $m\ell$ and integrate
once over $\xi$.  The result is:
\begin{equation}
  \label{aeqn3}
  \zeta(\xi) a(\xi) -\! {\beta_t\over\beta_l}\! \int_0^1 \!\! d\xi' 
  \Bigl[Z(\xi,\xi') + im\ell {\cal M}(\xi,\xi')\Bigr] {\cal 
    S}_{cS}[\xi']  =\xi^{3/2},
\end{equation}
where
\begin{equation}
  \label{Zdef}
  Z(\xi,\xi')=2\sqrt{\xi\over \xi'}-\ln\left|{\sqrt \xi' 
      +\sqrt \xi\over\sqrt \xi' -\sqrt \xi}\right|,
\end{equation}
and
\begin{equation}
  {\cal M}(\xi,\xi')=(\xi-\xi')\, 
  \ln\left|{\sqrt{\xi}+\sqrt{\xi'}\over\sqrt{\xi}-
      \sqrt{\xi'}}\right| +2\sqrt{\xi\xi'}-{2\xi\over 
    3}\sqrt{\xi\over\xi'}.
\end{equation}
The function ${\cal S}_{cS}[a(\xi)]$ is defined in (\ref{calS}).

Although they never wrote their equation in this form, CLN were
studying the equivalent of (\ref{aeqn3}) for the Dugdale model
(\ref{DD}) and $\eta=0$, that is, for ${\cal S}_{cS}[a(\xi)]=a(\xi)$,
when they concluded that this fully dissipationless system was
manifestly unstable. They reached this conclusion by noticing that, in
the Dugdale case as shown in (\ref{zetaDD}),
\begin{equation}
  \zeta(\xi)=\int_0^1 Z(\xi,\xi')\,d\xi'.
\end{equation}
Therefore, if one sets $m=v=0$ in (\ref{aeqn3}), so that
$\beta_t/\beta_l=1$, then $a=$ constant is trivially a null right
eigenfunction of the operator on the left-hand side (i.e.~it is
a non-trivial solution of the homogeneous equation).  Thus they
expected that $a(\xi)$ and, accordingly, the second term in $\Phi$ in
(\ref{Phidef}), must have a singularity near the origin in the
$m$-plane for small $v$.  By equating coefficients of $\xi^{3/2}$ on
both sides of (\ref{aeqn3}), they found
\begin{equation}
  a\cong {(3/4)\over (1-1/\kappa)v^2 -im\ell} 
\end{equation}
which implied a zero in $\Phi(m,v)$ (a pole in $\hat\chi_Y$) at $m\ell
= i(1-1/\kappa)v^2$. This unstable pole in the upper half $m$-plane
seemed to be consistent with the tip-stress analysis mentioned in
Section \ref{sec:intro} of the present paper.
 
What CLN failed to realize is that the left-hand side of (\ref{aeqn3})
is a singular integral operator that has null right eigenfunctions,
not just for $m=v=0$, but for a continuous range of non-zero values of
$m$ and $v$.  It is a well known property of singular integral
equations such as (\ref{aeqn3}) [see \cite{muskhelishvili}] that, like
inhomogeneous differential equations, they can have both homogeneous
and particular solutions.  That is precisely what happens here. The
CLN version of (\ref{aeqn3}) has no unique solution but, instead, has
a continuous, one-parameter family of solutions obtained from a
particular solution by adding an arbitrary amount of the null
(i.e.~homogeneous) solution.  We see no physical criterion for
selecting among these solutions, thus we conclude that this version of
the linear stability theory is mathematically ill-posed.

M.\ Marder \cite{marder-private} has pointed out to us that the
existence of the physically acceptable null eigenfunction of the
operator on the left-hand side of Eq.\ (\ref{aeqn3}) means that the
steady state solution whose stability we are studying may be
non-unique.  The null eigenfunction implies, at least in a linear
approximation, that the steady state solution can contain
symmetry-breaking, shear deformations in the absence of the perturbing
shear stress.  To compute the amplitudes of these deformations, and to
to find out precisely the conditions under which they exist, we would
have to carry out a fully nonlinear analysis of the zeroth-order
problem including shear crack opening displacements $U_S(x)$ as well
as the normal displacements $U_N(x).$  If this interpretation is
indeed correct, then what we have identified as a mathematical
pathology in the linear response calculation, is an equally
pathological lack of definition of the steady-state behavior of these
models. 

\begin{figure}[htbp]
  \centerline{\epsfxsize=3.3in \epsfbox{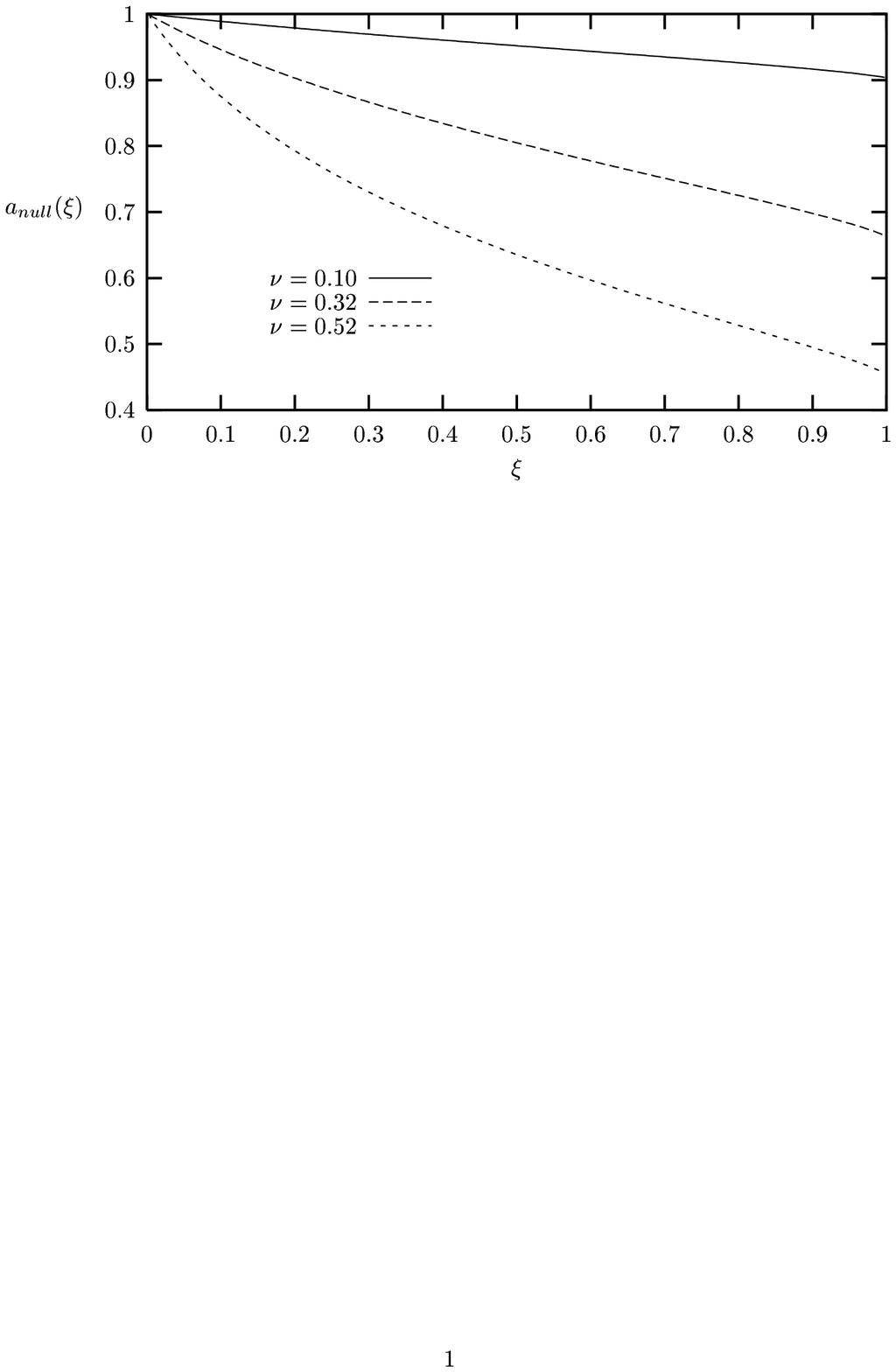}}
  \vspace{0.1in}
  FIG.~2. Well behaved right null eigenfunctions of 
  ${\cal L}_{\nu}$ for three different values of $\nu$.
  \label{fig:a_null}
\end{figure}
To illustrate the mathematical difficulties, we consider the special
case of (\ref{aeqn3}) with, as above, ${\cal S}_{cS}[a(\xi)]=a(\xi)$
and $m=0$.  This equation has the form
\begin{equation}
  \label{Lnu}
  {\cal L}_{\nu}*a(\xi)\equiv \zeta(\xi)a(\xi) - (1 - \nu) \! \int_0^1
  \!\! d\xi' Z(\xi,\xi') a(\xi') = \xi^{3/2},
\end{equation}
where $\nu=1-\beta_t/\beta_l\cong (1-1/\kappa)(v^2/2)$.  In
Fig.~\ref{fig:a_null}, we show null right eigenfunctions of ${\cal
L}_{\nu}$, i.e.~solutions of ${\cal L}_{\nu}*a_{null}(\xi)=0$, for
various values of $\nu$.  These functions $a_{null}(\xi)$ are all
perfectly well behaved throughout the interval $0\le\xi\le1$; there
seems to be no reason to reject them on either mathematical or
physical grounds.

The addition of dissipation changes this situation completely by
producing an unphysical singularity in the null eigenvectors and thus
a selection mechanism.  However, this does not turn out to be the
systematic and reliable selection mechanism that we had hoped to find.
As a first step in seeing what happens, we look at the same Dugdale
steady-state model, again with $m=0$, but now with a nonzero value of
the dissipation constant $\eta$.  That is, we study
\begin{equation}
  \label{aeqn4}
  {\cal L}_{\eta}*\tau(\xi)\equiv \zeta(\xi) a(\xi) - (1 - \nu)
  \!\int_0^1  \!\!d\xi' Z(\xi,\xi') \tau(\xi') = \xi^{3/2},
\end{equation}
where
\begin{equation}
  \tau(\xi)=a(\xi)+ {\eta v\over\ell}\,{da\over d\xi}.
\end{equation}
For numerical purposes, we find it best to eliminate $a(\xi)$ in favor
of $\tau(\xi)$ by writing
\begin{equation}
  \label{taua}
  a(\xi)={\ell\over\eta v}\int_0^{\xi} d\xi'\,\exp\left[{\ell\over\eta
      v}\,(\xi'- \xi)\right]\,\tau(\xi').
\end{equation}

Note that we have fixed $a(0)=0$.  Our argument for doing so is
analogous to one that Langer and Nakanishi \cite{LN} used in
discussing a model with a similar dissipative term.  For nonzero
$\eta$, we must avoid a discontinuous jump in the shear angle, and
therefore a delta-function spike in the shear cohesive stress at the
crack tip.  For $\eta v \ll \ell$, (\ref{taua}) produces a thin
boundary layer within which $a(\xi)$ rises rapidly from zero at $\xi =
0$.  Here is the first --- but unfortunately not the last --- occasion
in which we shall see a length scale much smaller than $\ell$ playing
a sensitive role in our analysis.  As mentioned in Section
\ref{sec:intro}, we do not believe that small-scale features of this
kind can be taken seriously as physically significant ingredients of
our theory; thus we discount results in this regime.  Nevertheless, we
must find out what happens mathematically.

\begin{figure}[htbp]
  \centerline{\epsfxsize=3in \epsfbox{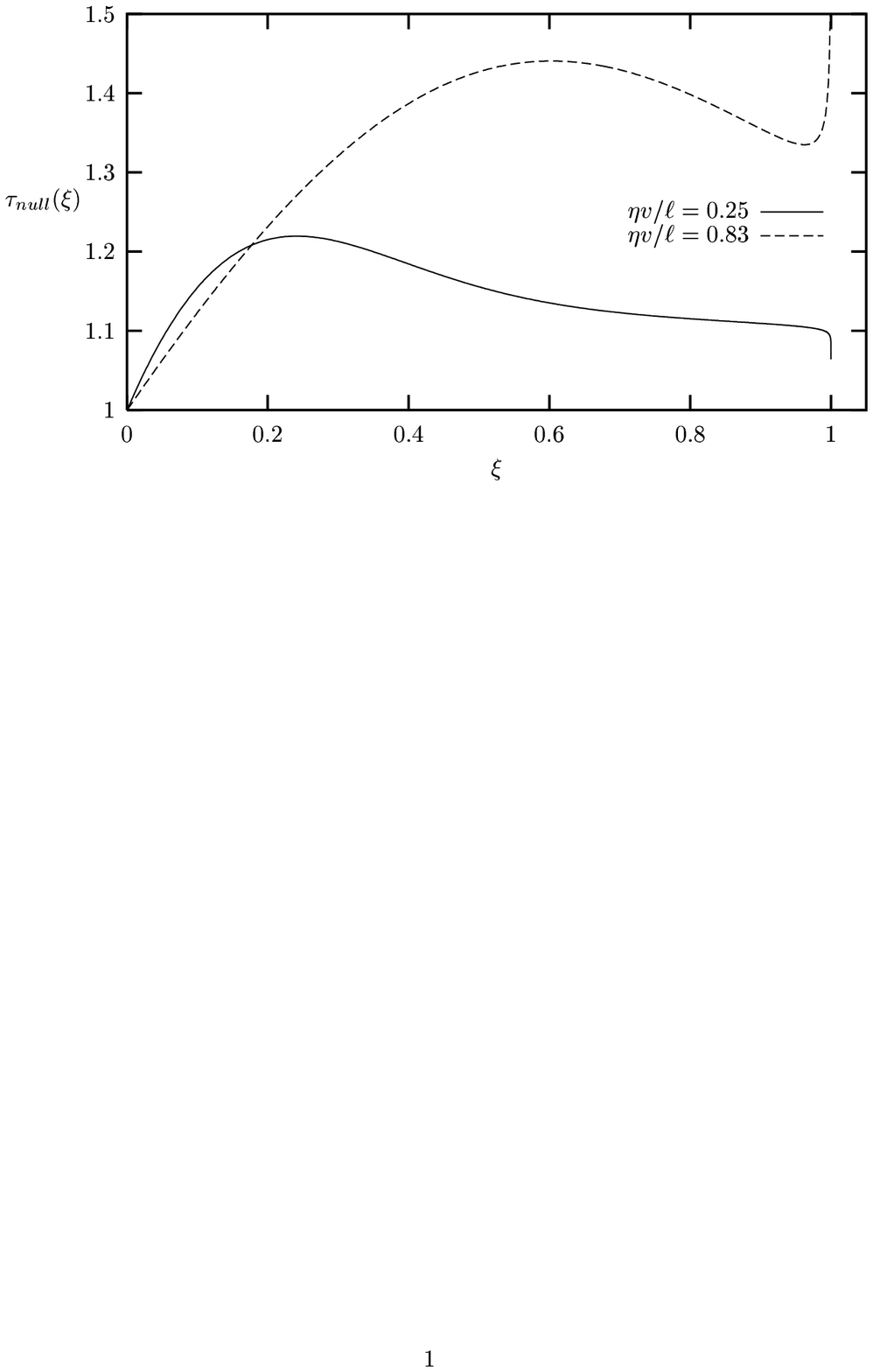}}
  \vspace{0.1in}
  FIG.~3. Right null eigenfunctions of ${\cal L}_{\eta}$ 
  for two values of speed $v$ and fixed $\eta=\delta$ (in our
  units $\eta$ is a length and is therefore measured in units of
  $\delta$ as all other lengths are).  We have normalized these
  functions to unity at $\xi=0$.  Note that the coefficient of the
  singularity at $\xi=1$ changes sign.
  \label{fig:tau_null}
\end{figure}
\begin{figure}[htbp]
  \centerline{\epsfxsize=3.3in \epsfbox{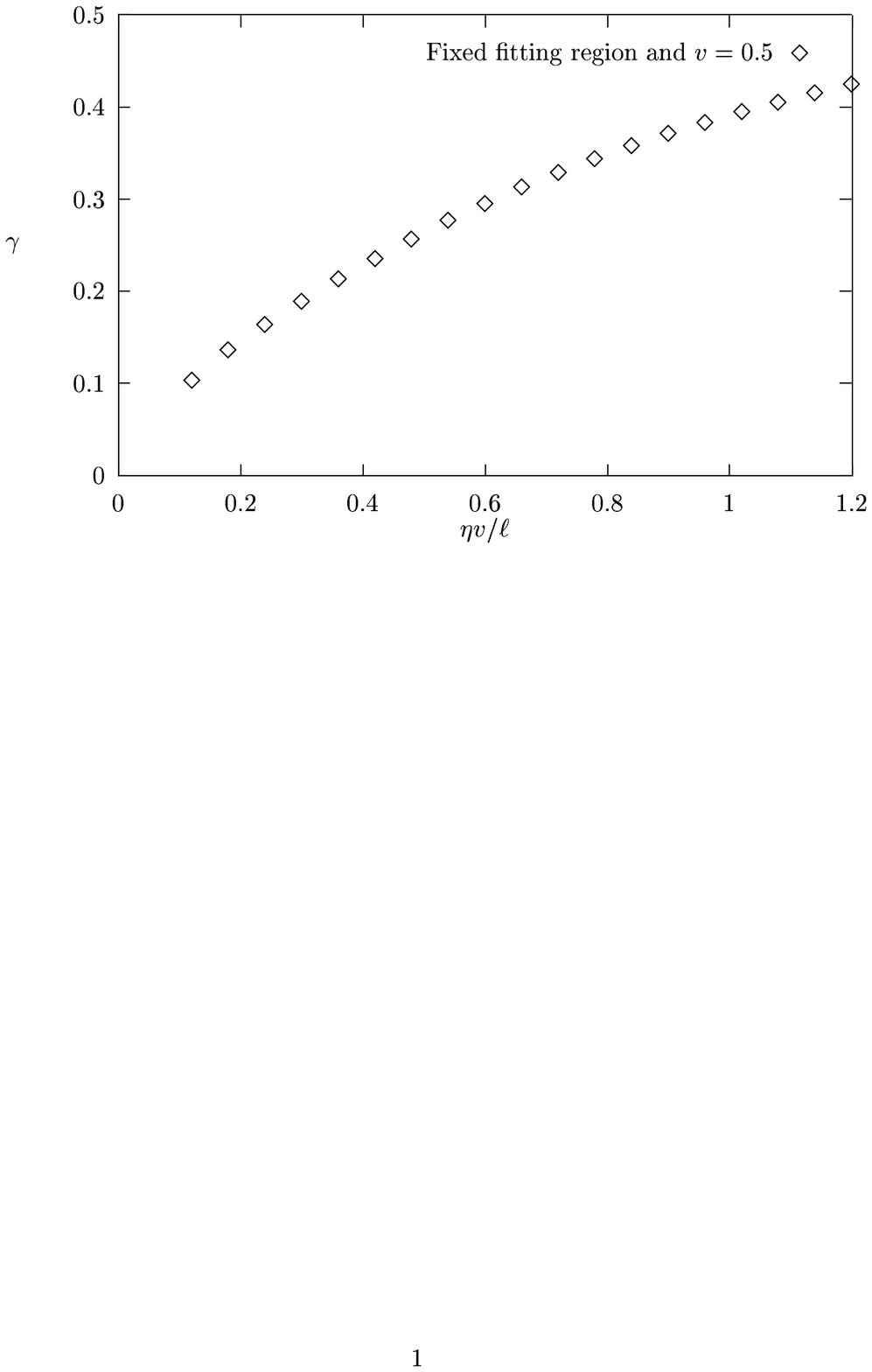}}
  \vspace{0.1in}
  FIG.~4. The exponents of the power law divergence of 
  the solutions at $\xi=1$ extracted by fitting the functional
  form of the discrete solution in an interval of fixed width near
  $\xi=1$.
  \label{fig:exponents}
\end{figure}

We have computed the null right eigenfunctions of ${\cal L}_{\eta}$,
say $\tau_{null}(\xi)$, and show several of them normalized to unity
at $\xi=0$ in Fig.~\ref{fig:tau_null}.  Most notably, addition of the
$\eta$ term in (\ref{aeqn4}) produces power-law singularities at
$\xi=1$ of the form $(1-\xi)^{-\gamma}$.  In Fig.~\ref{fig:exponents},
we show the exponent $\gamma$, determined by fitting the functional
form of $\tau_{null}$ to a power law in some region around $\xi=1$, as
a function of $\eta v /\ell$ for $\nu=0.1$. This singularity becomes
very weak at small values of $\eta v/\ell$ and therefore makes our
numerical analysis difficult in that region.  By using smaller and
smaller fitting regions, and correspondingly smaller numerical grid
spacings, we have estimated $\gamma$ for values of $\eta v/\ell$
somewhat less than those shown in the Figure.  We believe that
$\gamma$ remains positive all the way down to $\eta v /\ell = 0$,
implying that all of these null solutions for the stress are divergent
and therefore physically unacceptable.  The function $\gamma(\eta v
/\ell)$ seems not to depend sensitively on $\nu$.

\begin{figure}[htbp]
  \centerline{\epsfxsize=3in \epsfbox{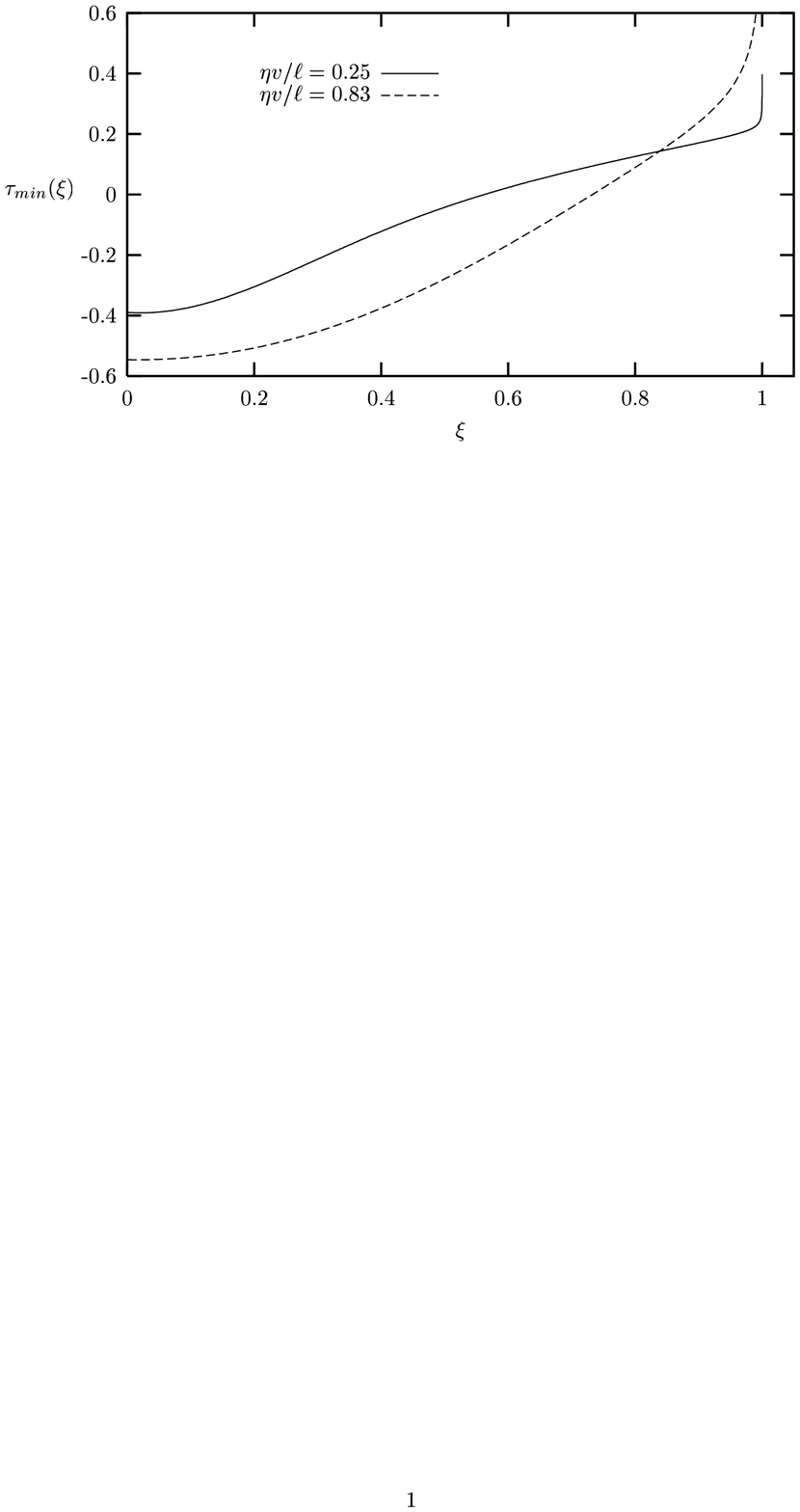}}
  \vspace{0.1in}
  FIG.~5.  The minimum norm solutions of the bending 
  response equation with the Dugdale normal cohesive stress.  The
  values of the parameters are identical to those in
  Fig.~\ref{fig:tau_null}.  Note that the coefficients of the
  singular part do not change sign.
  \label{fig:tau_min}
\end{figure}

We also have computed certain particular solutions of the
inhomogeneous equation (\ref{aeqn4}), the so-called ``minimum-norm''
solutions defined in Appendix C. We denote these by the symbol
$\tau_{min}(\xi)$, and show several of them in Fig.~\ref{fig:tau_min}
with parameters corresponding to the null solutions shown in
Fig.~\ref{fig:tau_null}.  These particular solutions also have
power-law singularities at $\xi=1$ with the same values of $\gamma$ as
those shown in Fig.~\ref{fig:exponents}.  The acceptable solutions of
(\ref{aeqn4}), therefore, have the form
\begin{equation}
  \tau(\xi)=\tau_{min}(\xi)+ C(v,\eta)\,\tau_{null}(\xi), 
\end{equation}
where the factor $C(v,\eta)$ must be chosen so that the singularities
in $\tau_{min}$ and $\tau_{null}$ cancel at $\xi=1$.

\begin{figure}[htbp]
  \centerline{\epsfxsize=3.3in \epsfbox{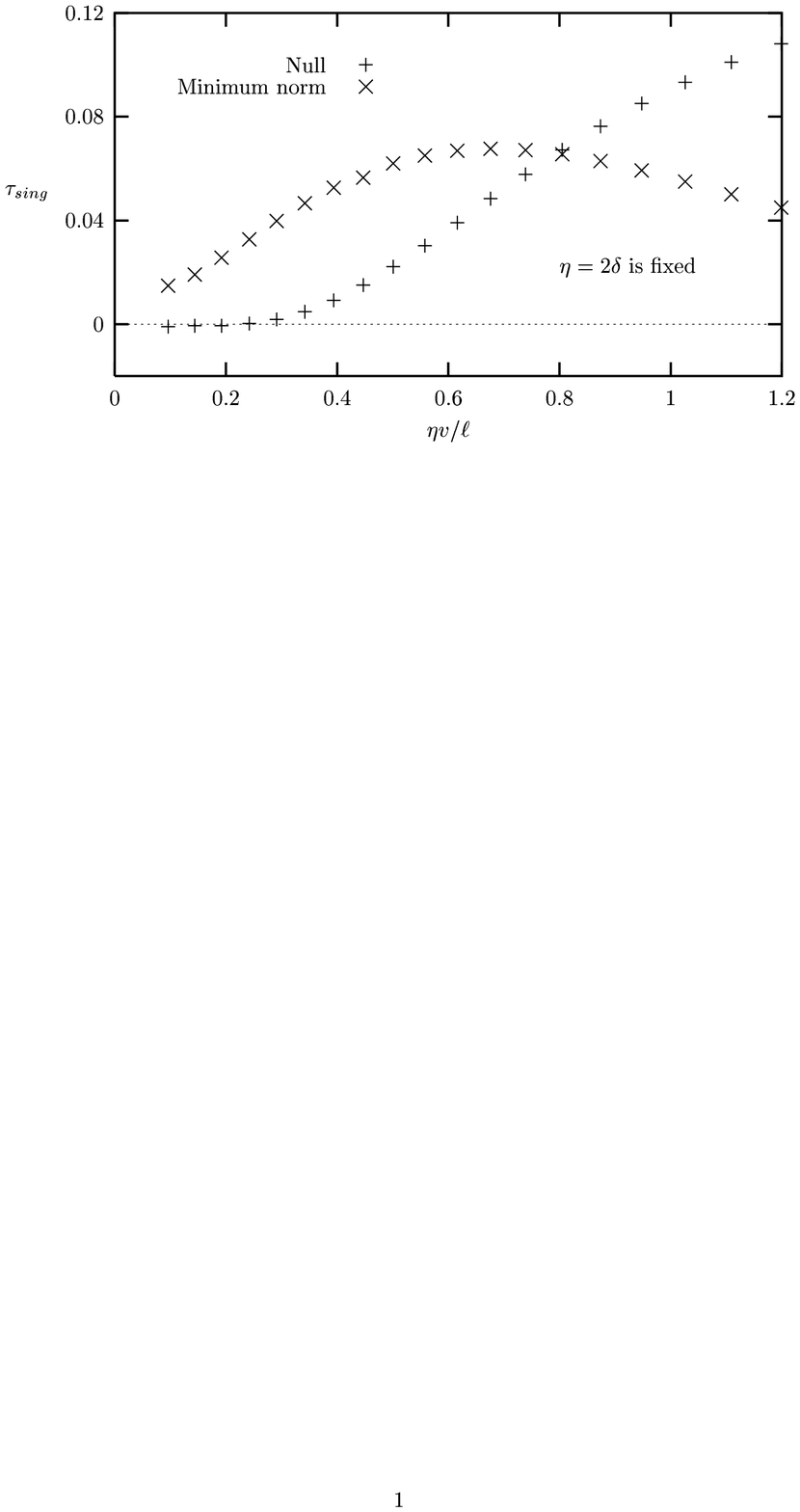}}
  \vspace{0.1in}
  FIG.~6.  The coefficients $\tau_{sing}=\lim_{\xi\to 1} 
  (1- \xi)^{\gamma}\tau(\xi)$ of the singular parts of the null
  and the minimum norm solutions for for a fixed $\eta=2\delta$.
  \label{fig:coefficients}
\end{figure}

The resulting solutions for $\tau(\xi)$ or, equivalently, $a(\xi)$
look much like those described by CLN near $v=0$.  The reason for this
behavior, as shown in Fig.~\ref{fig:coefficients}, is that the
coefficient of the divergent part of $\tau_{null}$, i.e.~$\tau_{sing}
\equiv \lim_{\xi\to 1}(1- \xi)^{\gamma}\,\tau_{null}(\xi)$, vanishes
at some value of $v$.  Therefore, $C(v,\eta)$ diverges at that value
of $v$, producing a divergence in $a(\xi)$.  At and near that value of
$v$, the selected solution $\tau(\xi)$ is dominated by the null
solution, which we know is very nearly a constant near $v=0$.
However, the fact that we recover something like the CLN solutions
near threshold does not mean that we also recover the CLN instability.
As we shall see, the results of the stability analysis are sensitively
dependent on the behavior of these weak singularities in the solutions
of the integral equation, and crude approximations do not seem to
work.

With the understanding that a nonzero dissipation coefficient $\eta$
may (or may not) determine a unique physically acceptable solution, we
return now to (\ref{aeqn3}) and report results of numerical solutions
with nonzero $m$ and various choices of the normal cohesive stress as
described in Section \ref{sec:steady}.  In general, once we have found
an acceptable set of solutions for $a(\xi)$ for various values of $m$
and $v$, we can evaluate $\hat\chi_Y^{-1}(m,v)$ in (\ref{chi-scaled})
and look for poles in the complex $m$-plane.  Because $mW$ must be
indefinitely large, we can neglect the first term on the right-hand
side of (\ref{chi-scaled}) (the ``$T$ stress'') and consider only
$\Phi(m,v)$ defined in (\ref{Phidef}).  Remember that the zeroes of
$\Phi(m,v)$ then correspond to the poles of $\hat\chi_Y(m,v)$, and
that a change in stability occurs whenever a pole crosses the real $m$
axis.  We emphasize again that, because of the weak singularities in
$a(\xi)$ at $\xi=1$, these calculations become numerically difficult
when $\eta v/\ell$ becomes appreciably smaller than unity.  This
numerically troublesome region is also where we are least confident
about the physical basis for our cohesive-zone models.

\begin{figure}[htbp]
  \centerline{\epsfxsize=3.3in \epsfbox{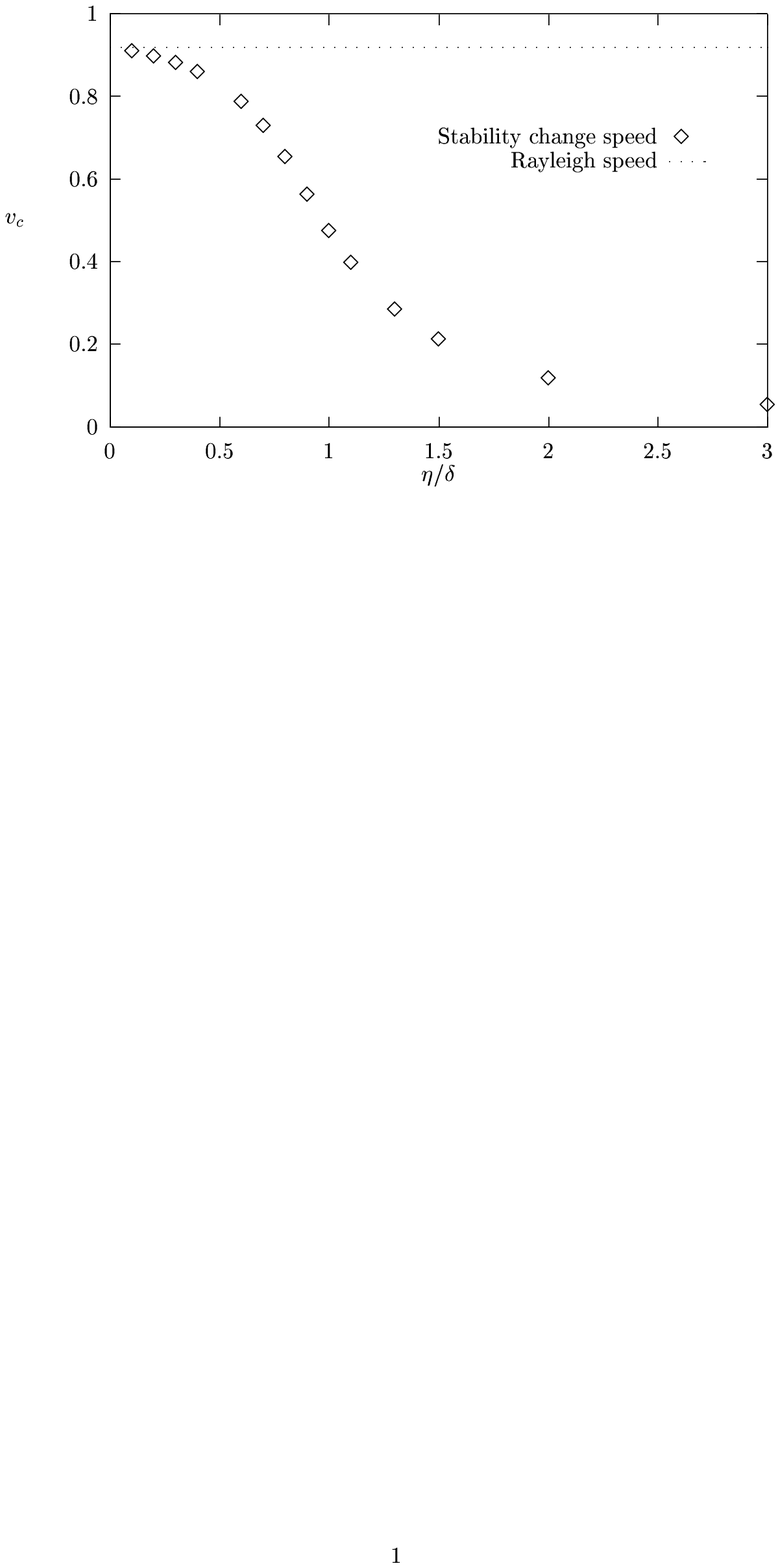}}
  \vspace{0.1in}
  FIG.~9.  The critical velocity $v_c$ as a function of the 
  shear dissipation coefficient $\eta$ for the Dugdale normal
  cohesive stress.
  \label{fig:threshold}
\end{figure}

The simplest non-trivial case is the Dugdale model for the normal
cohesive stress, i.e.~Eq.~(\ref{DD}), $\sigma_{cN}[\zeta(\xi)]=1$, but
with nonzero $\eta$ in the cohesive shear stress factor $\tau_{cS}
[a(\xi)]$ defined in (\ref{taudef}).  This is the case in which we
thought the stabilizing effect of $\eta$ should become apparent, and
perhaps it does.  Indeed, we now find stability at small crack speeds.
\begin{figure}[htbp]
  \centerline{\epsfxsize=3.3in \epsfbox{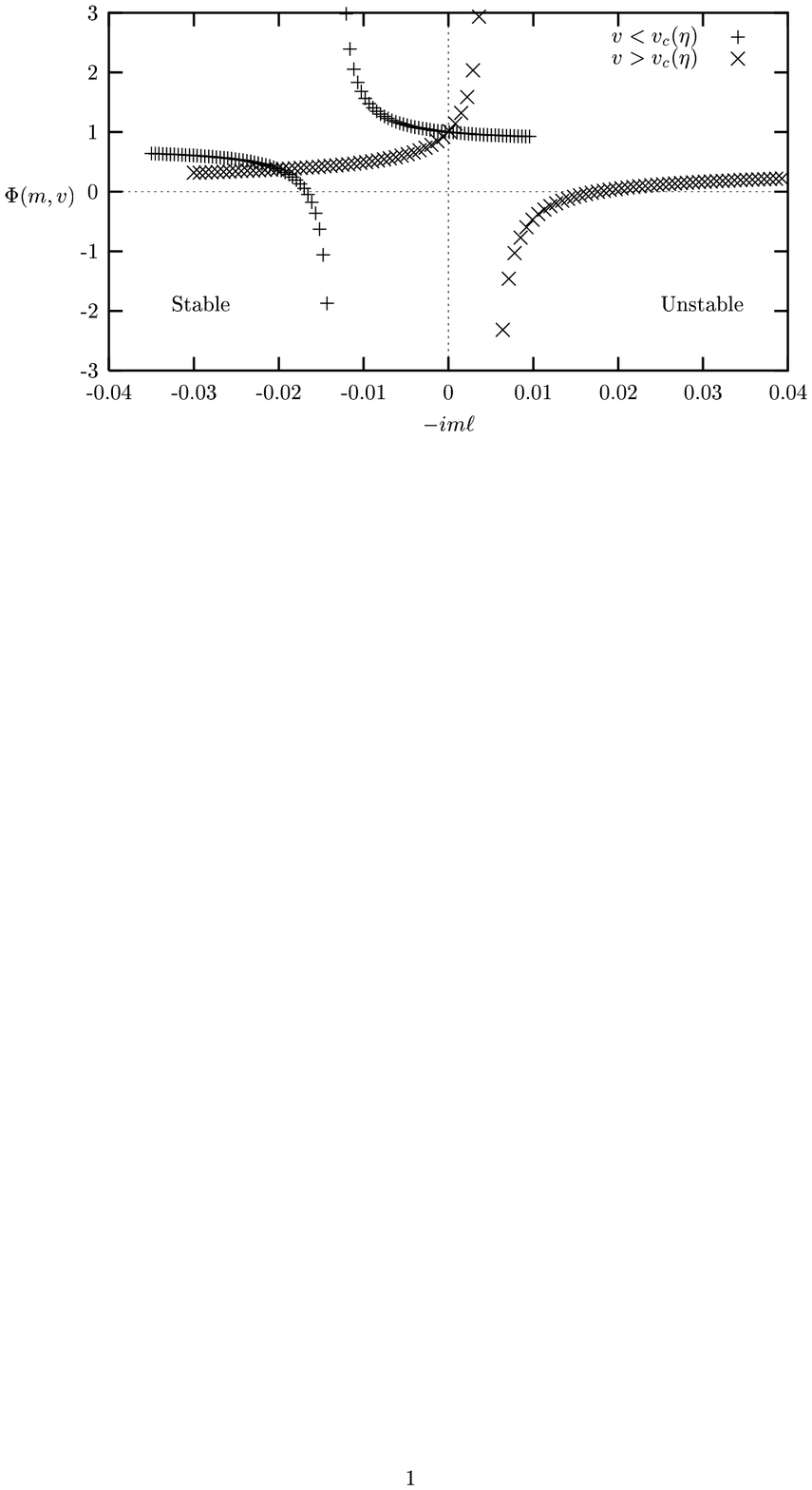}}
  \vspace{0.1in}
  FIG.~7.  The dependence of $\Phi(m,v)$ on $-im\ell$ for
  $\eta=\delta$ and two values of the crack's speed, one above and
  the other below the critical speed $v_c(\eta).$
  \label{fig:Phi}
\end{figure}
\begin{figure}[htbp]
  \centerline{\epsfxsize=3in \epsfbox{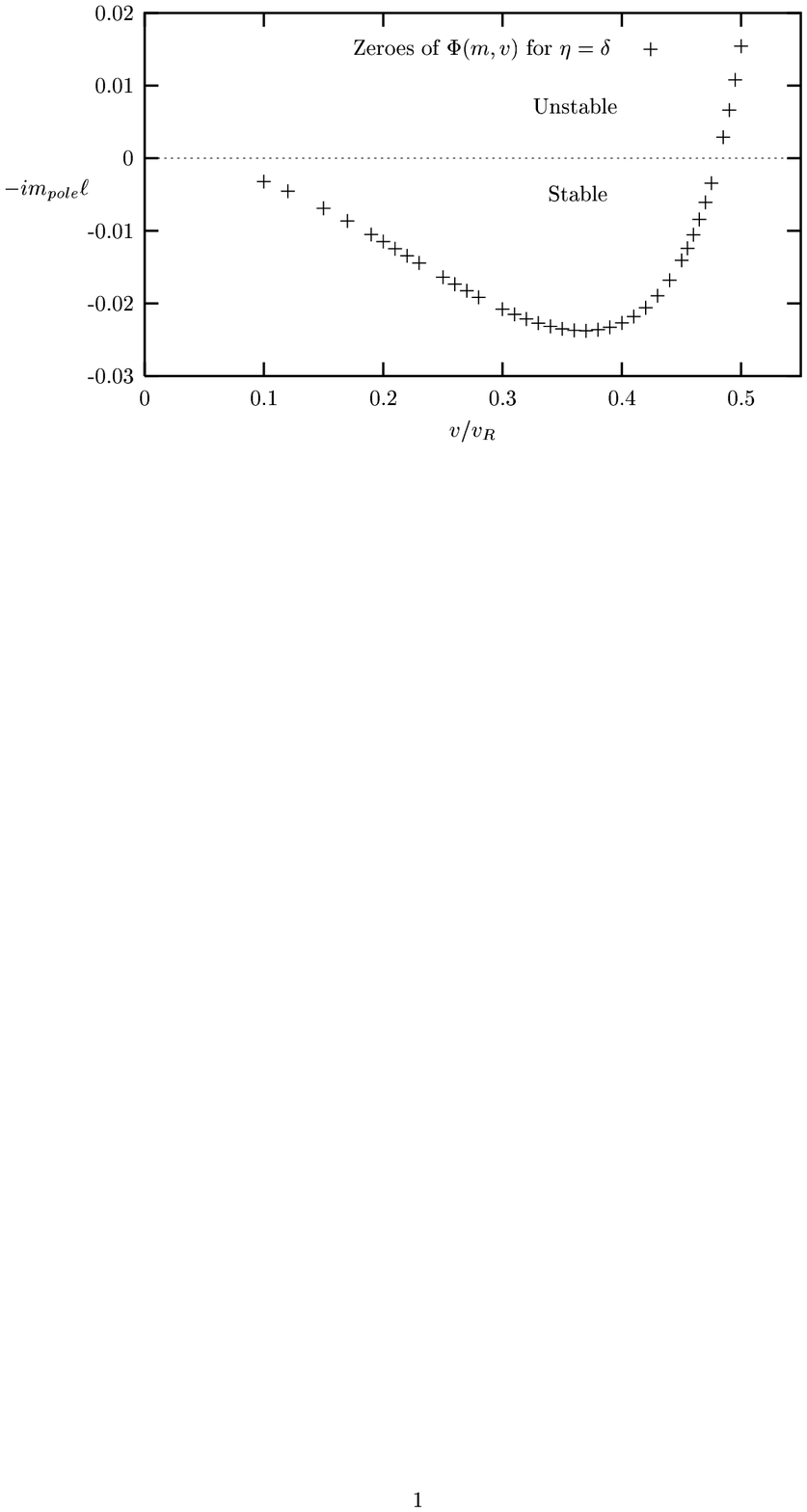}}
  FIG.~8.  Zeros of $\Phi(m,v)$ in the complex $m$-plane plotted as
  $-im_{pole}\ell$ vs.~$v/v_R$ for Dugdale normal cohesive
  stress.  $\eta = \delta$ is fixed.  These zeroes are located on
  the imaginary-$m$ axis.
  \label{fig:dugdale_poles}
\end{figure}
Recall from our analysis of (\ref{aeqn4}) that the physically
acceptable solution for $a(\xi)$ with $m=0$ diverges at a non-zero
value of the velocity, say $v_c(\eta)$.  As a result, the behavior of
$\Phi(m,v)$ as a function of $-im\ell$ is radically different for
values of the crack speed above and below this critical speed.  It is
clear from Fig.~\ref{fig:Phi} that, for $v > v_c$, the function
$\Phi(m,v)$ has a zero in the unstable half of the $m$-plane, and that
the system changes from stable to unstable as $v$ increases through
$v_c$.  We plot the position of this zero of $\Phi(m,v)$,
i.e.~$-im_{pole}\ell$, as a function of crack speed in
Fig.~\ref{fig:dugdale_poles}.  The function $v_c(\eta)$ is shown in
Fig.~\ref{fig:threshold}.  Note that $v_c$ decreases with increasing
$\eta$, which is perhaps not surprising.  The parameter $\eta$ occurs
in our equations only in the combination $\eta v/\ell$, and this term
should provide the dominant $v$-dependence at small speeds. At larger
speeds, of course, the ``relativistic'' $v^2$ dependence of the other
terms in $\Phi(m,v)$ start playing important roles and presumably are
responsible for the transition to instability.
Because this stability transition occurs at $m=0$, the interesting
behavior takes place at values of $|m\ell|$ that are small enough to
be consistent with our approximations.  Thus we are forced to the
conclusion that increasing the dissipation strength $\eta$ decreases
near-threshold stability in this model.  However, the ratio $\eta
v/\ell$ is small everywhere in Fig.~\ref{fig:threshold}, and actually
decreases as $\eta$ increases.  Thus, this apparently unphysical
result occurs only in the region where the dissipation length $\eta v$
is small compared to the size of the cohesive zone $\ell$.  If we
simply ignore this small-$v$ region and look only at situations where
$\eta v\ge\ell$, then we recover something like the CLN instability,
i.e.~the large-$v$ behavior shown in Fig.~\ref{fig:dugdale_poles}.
But the model then gives us no sensible description of the transition
from stability to instability at small velocities.
\begin{figure}[htbp]
  \centerline{\epsfxsize=3.3in \epsfbox{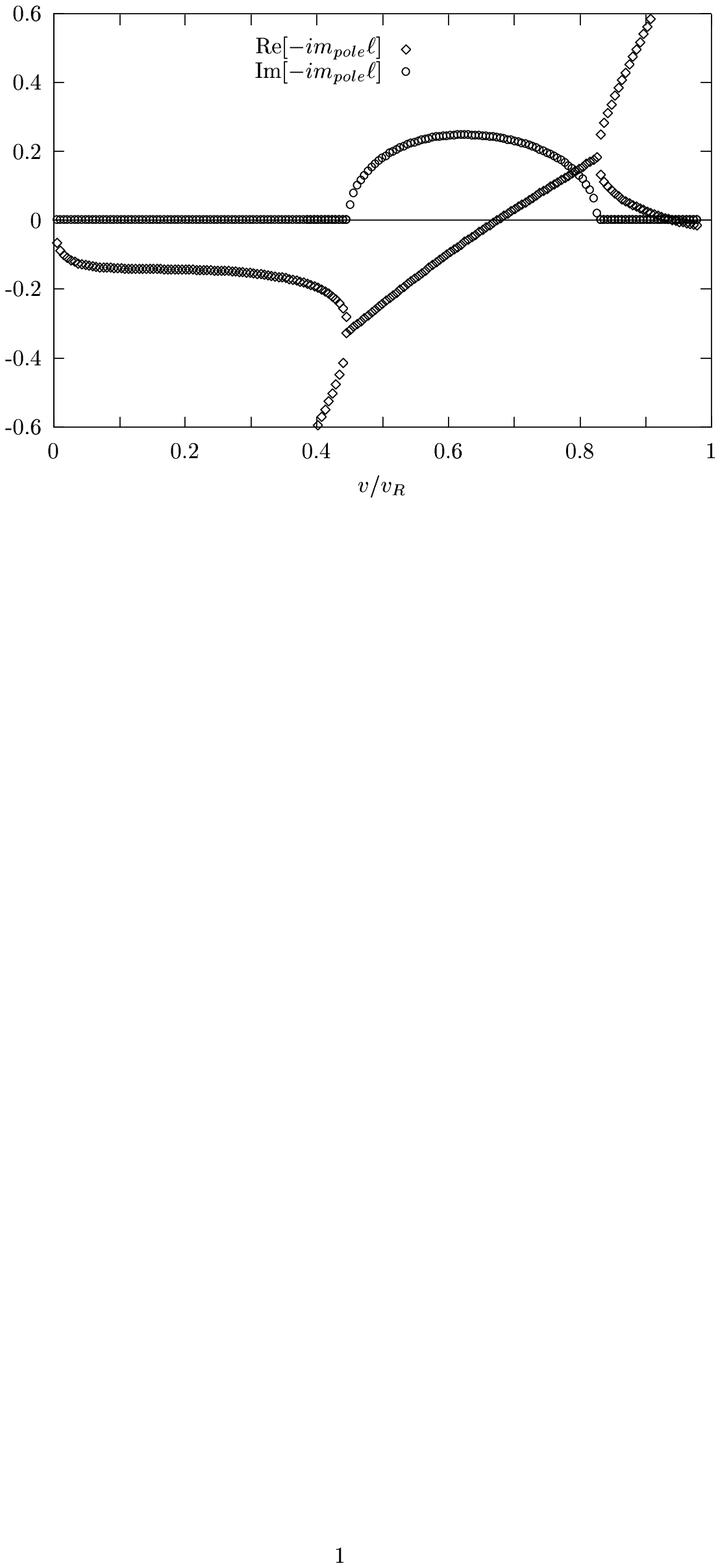}}
  \vspace{0.1in}
  FIG.~10.  The real and imaginary parts of the zeros
  $-im_{pole}\ell$ of $\Phi(m,v)$ in the complex $m$-plane for the
  model with rate-saturating dissipation in the normal cohesive
  stress.  The values of the various dissipation coefficients are
  fixed at $\eta=\delta$, $\lambda_1\Sigma_0 = 1$ and
  $\lambda_2\Sigma_0 = 0.05$.
  \label{fig:lambda_poles1}
\end{figure}
The next case that we examined produced our biggest surprise and our
most convincing evidence that this class of models is deficient in
important ways.  The normal cohesive stress shown in
(\ref{continuous}), for which the non-dissipative factor vanishes
continuously at the back edge of the cohesive zone, appears to us to
provide a perfectly reasonable model of forces acting at a crack tip.
However, when we tried to use (\ref{continuous}) in the integral
equation (\ref{aeqn3}), we found a null eigenfunction of of the
operator on the left-hand side that, so far as we can tell, has no
unphysical singularity anywhere in the interval $0\le\xi\le 1$.  [We
set $a(0)=0$ as in (\ref{taua}) and solved (\ref{aeqn3}) for
$\tau_{cS}(\xi)$.]  We conclude, therefore, that --- just as in the
much simpler case studied by CLN --- this model is mathematically
ill-posed for purposes of stability analysis.

\begin{figure}[htbp]
  \centerline{\epsfxsize=3.3in \epsfbox{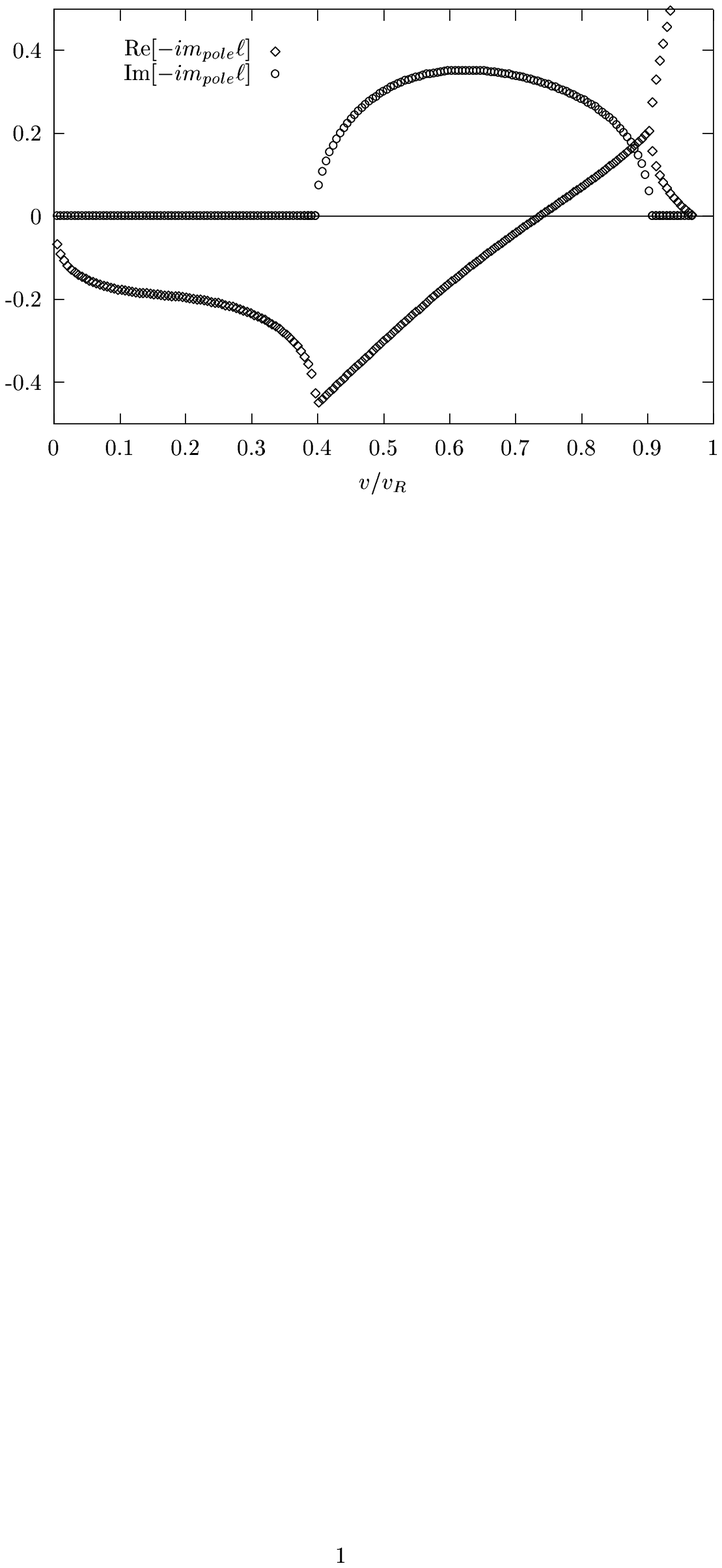}}
  \vspace{0.1in}
  FIG.~13.  Same as for Fig.~\ref{fig:lambda_poles1} except that
  $\lambda_1\Sigma_0 = 2$ and $\lambda_2\Sigma_0 = 0.1$ so that
  their ratio is kept constant.
  \label{fig:lambda_poles3}
\end{figure}

We do not claim to understand what it means that an apparently
well-posed physical model of dynamic fracture has a mathematically
undefined response to infinitesimally small bending perturbations. Nor
do we know whether we have accidentally found just one unusually
pathological example, or whether many such models behave in this way
--- perhaps all models with cohesive stresses that decrease
continuously to zero at large separations between the crack faces.  In
any case, this example diminishes our confidence in this approach to
fracture dynamics.
\begin{figure}[htbp]
  \centerline{\epsfxsize=3.3in \epsfbox{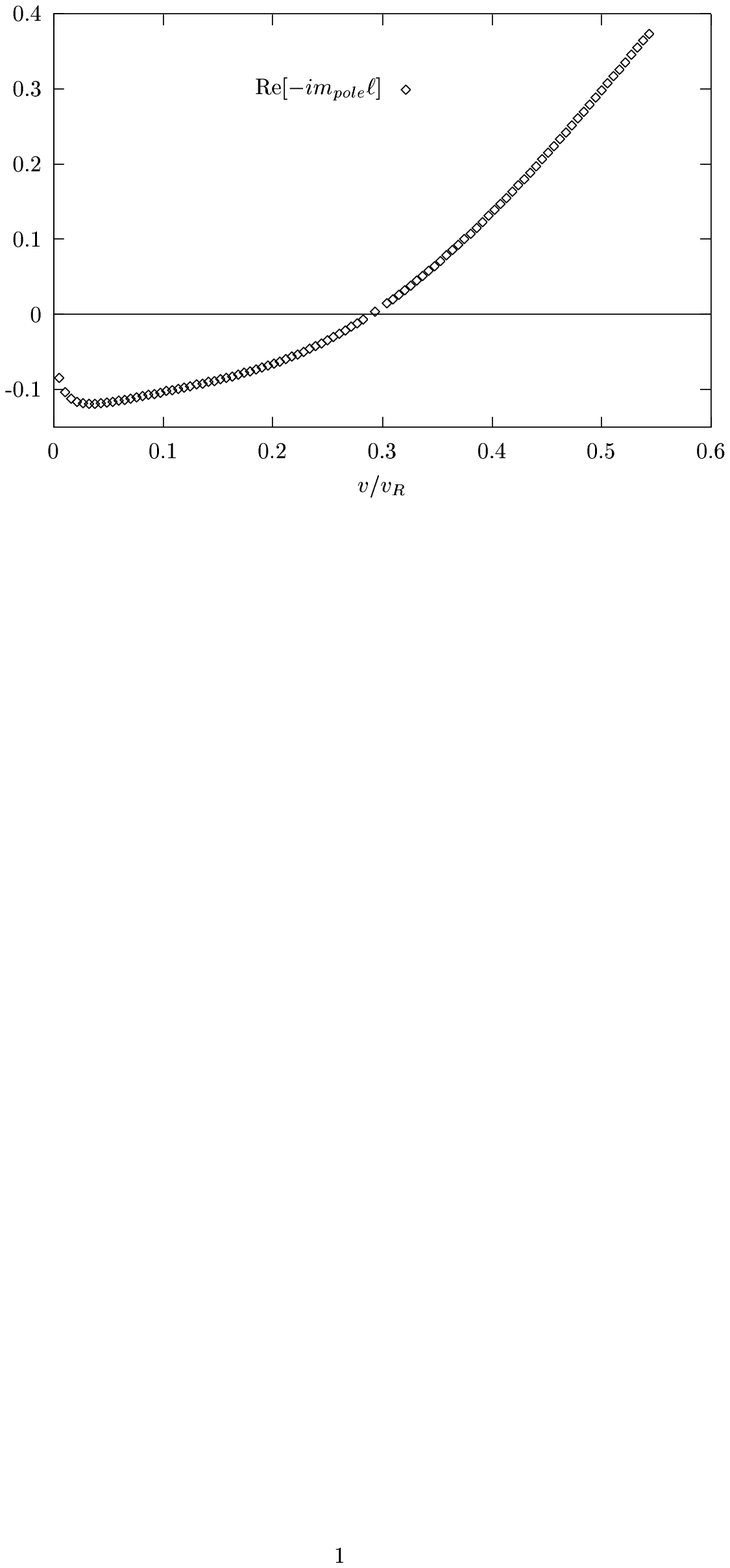}}
  \vspace{0.1in}
  FIG.~12.  Same as for Fig.~\ref{fig:lambda_poles1} except 
  that $\eta = 2\delta$.  The imaginary part of the
  $-im_{pole}\ell$ is not plotted since it vanishes identically
  for these values of the parameters.
  \label{fig:lambda_poles2}
\end{figure}
The one class of models for which we have found somewhat sensible
behavior is that for which the normal cohesive forces saturate at high
opening rates, i.e.~the model shown in (\ref{rate-saturate}).  Just as
in the Dugdale model discussed previously, divergences at the back end
of the cohesive zone provide a selection criterion, thus the model
seems to be mathematically well posed.  We have found that adding
dissipation to the normal cohesive stress changes some but not all
aspects of the behavior of the poles of $\hat\chi_Y(m,v)$ while
keeping them in the region of the complex $m$-plane where $|m\ell|$ is
small enough to be consistent with our approximations.  Some typical
results are shown in Figs.~\ref{fig:lambda_poles1},
\ref{fig:lambda_poles2} and \ref{fig:lambda_poles3}, where we plot the
real and imaginary parts of $-im_{pole}\ell$ as functions of $v$ for
two different values of $\eta=\delta$ and $\eta=2\delta$, and two sets
of values of $\lambda_{1,2}$ chosen in such a way as to keep their
ratio constant.  As before, we find stability at small $v$ and a
transition to instability at a critical velocity $v_c(\eta, \lambda_1,
\lambda_2)$.  The values of $\eta v/\ell$ at the transition are
$0.651$, $0.510$, and $0.513$ respectively.  Thus we again seem to be
in the physically implausible region in which the microscopic length
$\ell$ controls the behavior.  For the smaller value of $\eta$, the
transition is a Hopf bifurcation which determines a characteristic
wavenumber, i.e.~a nonzero real part of $m$, at which the initial
instability takes place.  Increasing dissipation in the normal
cohesive stress increases the critical speed thus making the crack
more stable.  For the larger value of $\eta$, the transition occurs at
$m=0$ and, ominously, at a smaller value of $v_c$.  Thus once again we
have a situation in which increasing $\eta$, i.e.~increasing the
resistance to shear deformations, reduces rather than increases the
threshold for instability.

\subsection*{In summary, our results are the following:}

The CLN model, which is a simple Dugdale cohesive-zone model with no
dissipation in either the normal or shear stresses, is not
mathematically well posed for purposes of studying stability of mode-I
fracture against out-of-plane bending perturbations.

Similarly, a more complicated and apparently realistic model that
contains both normal and shear dissipation, and for which the normal
stress vanishes continuously at the back end of the cohesive zone, is
mathematically ill-posed. The stability analysis does not make sense
for the same mathematical reasons that caused the CLN analysis to
fail.

Addition of a simple linear dissipative shear stress converts the
Dugdale model into a mathematically well defined system that is stable
at the Griffith threshold and undergoes an instability at a higher
velocity.  However, this behavior depends sensitively on the
mathematical structure of the cohesive zone at unphysically small
length scales, and therefore seems to depend on artificial features of
the model.  One symptom of the artificiality of the model is that the
critical velocity at which the crack becomes unstable decreases with
increasing dissipation strength.

Our fourth, most complex, model, which contains both shear dissipation
and a nonlinear dissipative contribution to the normal stress, is
mathematically well posed and exhibits physically reasonable behavior
at least in some regions of its parameter space.  However, like the
dissipative Dugdale model, it also loses stability with increasing
strength of the shear dissipation.  We have no reason to believe that
this model is representative of any broad class of physically
motivated fracture models, or that its behavior is a robust feature of
such a class.

\section*{Acknowledgements}

This research was supported by DOE Grant No. DE-FG03-84ER45108 and
also in part by NSF Grant No. PHY94-07194.

\bibliographystyle{prsty}
\bibliography{fracture}

\appendix

\section{Derivation of the steady-state equations}
\label{sec:zeroapp}

The Wiener-Hopf equations for the steady-state theory are most easily
written in a Fourier representation such that, for example,
\begin{mathletters}
  \begin{equation}
    \Sigma_{N}(x) =  \int_{-\infty}^{+\infty} {dk\over 2\pi}\, 
    \hat\Sigma_N(k)\,e^{ikx},
  \end{equation}
  \vspace{-0.2in}
  \begin{equation}
    \hat\Sigma_{N}(k) = \int_{-\infty}^{+\infty}dx\,
    \Sigma_{N}(x)\,e^{-ikx}.
  \end{equation}
\end{mathletters}
In this notation, the zeroth order equation [CLN(3.9)] is
\begin{eqnarray}
  \nonumber
  \hat\Sigma_{N}(k)&=&\hat\Sigma_{cN}^{(+)}(k) +
  \hat\Sigma_{N}^{(-)}(k)\\ &=&2\pi\Sigma_{N\infty}\delta(k) 
  -\hat F(k)\,\hat{U}_{N}^{(+)}(k)\label{WH0}
\end{eqnarray}
The superscripts in parentheses $(\pm)$ indicate functions that have
singularities only in the upper (or lower) half $k$-planes because
they are Fourier transforms of functions that are nonzero only on the
positive (or negative) $x$-axis. Thus, the first line of (\ref{WH0})
tells us in Fourier language that the stress $\Sigma_N(x)$ consists of
a part $\Sigma_N^{(-)}(x)$ that is nonzero only ahead of the crack,
$x<0$, plus the normal component of the cohesive stress,
$\Sigma_{cN}(x)$, which is nonzero for $x>0$ within the cohesive zone.
The second line of (\ref{WH0}) relates these stresses to $U_N(x)$,
which is nonzero only behind the crack tip, $x>0$. The quantities
$\Sigma_{N\infty}$ and $\kappa$ are defined in Section \ref{sec:zero}.

The Wiener-Hopf kernel $\hat F(k)$ is
\begin{equation}
  \label{F}
  \hat F(k) = \left[\left(\kappa\over 2W \right)^2 + b^2(v) k^2
  \right]^{1/2} \equiv b(v)\,(\alpha^2+k^2)^{1/2}.
\end{equation}
This expression is a simple interpolation between exact results in the
limits $kW\to\infty$ (the only case of direct interest here) and
$kW\to 0$. The quantity 
\begin{equation} 
  \alpha\equiv{\kappa\over 2\,b(v)\,W}
\end{equation}
is infinitesimally small in the limit of very large $W$. In addition
to producing a correct value for $\hat F(0)$, its presence here
defines the analytic properties of $\hat F(k)$ in the complex $k$-
plane.

To solve the Wiener-Hopf equation, (\ref{WH0}), we first factor the
kernel (\ref{F}): 
\begin{equation} 
  \hat F(k)=\hat F^{(+)}(k)\,\hat F^{(-)}(k),
\end{equation}
with
\begin{equation}
  \label{Ffactors}
  \hat F^{(+)}(k)=b(v)\,(\alpha+ik)^{1/2};~~~\hat F^{(-)}(k) =
  (\alpha-ik)^{1/2}.
\end{equation}
In terms of these factors, the solutions for the displacement and
stress fields are, respectively:
\begin{equation}
  ik\,\hat U_N^{(+)}(k)={1\over \hat F^{(+)}(k)} 
  \,\left[{\Sigma_{N\infty}\over \hat F^{(-)}(0)} -
    ik\,\hat\Lambda_{cN}^{(+)}(k)\right];\label{UN}
\end{equation}
and
\begin{equation}
  ik\,\hat\Sigma_N^{(-)}(k)=-\hat F^{(-)}(k)\, 
  \left[{\Sigma_{N\infty}\over \hat F^{(-)}(0)} 
    +ik\,\hat\Lambda_{cN}^{(-)}(k)\right];\label{SigmaN}
\end{equation}
where
\begin{equation}
  \hat\Lambda_{cN}^{(\pm)}(k)=\mp\int{dk'\over 2\pi i}\, {1 \over k'-
    k \pm i\epsilon}\,{\hat\Sigma_{cN}^{(+)}(k') \over \hat
    F^{(-)}(k')}.
\end{equation}

A quick way to obtain the condition for nonsingular stress, i.e.~the
Barenblatt condition, is the following.  In (\ref{SigmaN}), the factor
$\hat F^{(-)}(k)\sim k^{1/2}$ for large $k$ produces the usual
$|x|^{-1/2}$ singularity in the stress at small $|x|$.  Both terms
inside the square brackets in (\ref{SigmaN}) are constants at large
$k$; thus the Barenblatt condition is simply the requirement that
these constants cancel each other. Specifically, 
\begin{eqnarray}
  \label{BCondapp}
  \Sigma_{N\infty} & = & \hat F^{(-)}(0)\,\int {dk\over 2\pi}\, 
  {\hat\Sigma_{cN}^{(+)}(k)\over \hat F^{(-)}(k)} \cr
  & = & \left[\kappa\over 2\pi\, b(v)\, W\right]^{1/2} \int_0^{\ell}
  {dx\over\sqrt{x}}\, \Sigma_{cN}[U_{N}(x)].
\end{eqnarray}
The second form of this result is the same as that shown in
(\ref{BCond}). We have redundantly inserted the length of the cohesive
zone, $\ell$, as the upper limit of integration as a reminder that
$\Sigma_{cN}[U_{N}(x)]$ vanishes for larger values of $x$.

An advantage of this technique is that we can substitute
(\ref{BCondapp}) into (\ref{UN}) to obtain an expression for $\hat
U_N^{(+)}(k)$ from which the singularity has been eliminated:
\begin{equation}
  ik\,\hat U_N^{(+)}={1\over \hat F^{(+)}(k)}\, \int{dk'\over
    2\pi}\,{k'\over k'-k+i\epsilon} \,{\hat\Sigma_{cN}^{(+)}(k')\over
    \hat F^{(-)}(k')}.\label{ikU}
\end{equation}
It is now safe and useful to compute the Fourier transform of
(\ref{ikU}) and, in doing so, to take the limit $W\to\infty$,
i.e.~$\alpha\to 0$ in the factors $\hat F^{(\pm)}(k)$.  The result is
Eq.~(\ref{U'x}).

\section{Derivation of the first-order equations}
\label{sec:firstapp}

Our starting point is the first-order equation [CLN(5.6)] that relates
the shear components of the displacement and stress:
\begin{eqnarray}
  \nonumber  
  &&\hat\Sigma_{S}(k,m) = \hat\Sigma_{cS}^{(+)}(k,m) +
  \hat\Sigma_{S}^{(-)}(k,m) = 2\pi \hat E_m\delta(k-m) \\
  && -\,\hat G(k,m)\,\hat U_{S}^{(+)}(k)+\hat L(k,m)\,\hat 
  U_{N}^{(+)}(k-m)\,\Ym, \label{WH1}
\end{eqnarray}
In analogy with (\ref{WH0}), $\hat\Sigma_{S}^{(-)}(k,m)$ is the
Fourier transform of the shear stress in the unbroken region $x<0$,
and $\hat\Sigma_{cS}^{(+)}(k,m)$ is the Fourier transform of the
cohesive shear stress.

The various ingredients of (\ref{WH1}) are:
\begin{equation}
  \hat E_m\equiv \epm+im \Delep\Ym;~~~\Delep=\epinfn - \epinft.
\end{equation}
The function which plays the role of the Wiener-Hopf kernel, the
analog of $\hat F(k)$ in (\ref{WH0}), is
\begin{equation}
  \hat G(k,m)={2\over q_tv^2(k-m)^2}\Bigl(k^2q_lq_t-
  q_0^4\Bigr),\label{G}
\end{equation}
where
\begin{eqnarray}
  q_l^2 & \equiv & k^2-{v^2\over\kappa}(k-m)^2, \cr \vspace{0.1in}
  q_t^2 & \equiv &k^2-v^2(k-m)^2, \cr q_0^2 & \equiv & k^2-{v^2\over
  2}(k-m)^2.
\end{eqnarray}
Finally, the function which determines the coupling between the
oscillating shear stress and the unperturbed crack is:
\begin{eqnarray}
  && \hat L(k,m) = {2\,ikq_l\over v^2}
  \Bigl[\Bigl({m\over k-m}\Bigr)^2-\beta_0^2\Bigr] \\ \nonumber
  && + {2\,ikq_0^2\over q_tv^2}
  \Bigl[1-\Bigl({m\over k-m}\Bigr)^2 -v^2\Bigl({k-m\over 
    k}\Bigr)\Bigr] \\ \nonumber
  && + \Bigl({2\,i\over v^2}\Bigr)\,|k-m| \Bigl[{\beta_0^2\over 
    \beta_l} \Bigl(\beta_l^2k +m\Bigr) -\beta_t \Bigl(\beta_0^2k + m +
  {mv^2\over 2}\Bigr)\Bigr]
  \label{Ldef}
\end{eqnarray}
Despite appearances, $\hat G(k,m)$ and $\hat L(k,m)$ are finite at
$k=m$ and $v=0$.

We now write the Wiener-Hopf kernel (\ref{G}) in the form:
\begin{equation} 
  \hat G(k,m)=\hat G^{(+)}(k,m)\,\hat G^{(-)}(k,m),
\end{equation}
where the superscripts $(\pm)$ have their usual significance. CLN find
these factors to be: 
\begin{equation}
  \hat G^{(+)}(k)={\beta_l\,b(v)\over\beta_t}\,{i(k-\krp)
    \over[\epsilon+i(k-\ktp)]^{1/2}}\,\exp[N^{(+)}(k)],
\end{equation}
and
\begin{equation}
  \hat G^{(-)}(k)={-i(k-\krm) \over[\epsilon-i(k-\ktm)]^{1/2}} 
  \,\exp[N^{(-)}(k)];
\end{equation}
where
\begin{eqnarray}
  N^{(+)}(k) & = &  -\int_{\klp}^{\ktp}\,{dp\over\pi}\,{\varphi(p)\over
    p-k}, \cr
  N^{(-)}(k) & = &\int_{\ktm}^{\klm}\, {dp \over \pi}\,
  {\varphi(p)\over p - k}, \\
  \tan\varphi(p) & = & {p^2|q_l(p)\,q_t(p)|\over q_0^4(p)};
\end{eqnarray}
and
\begin{equation}
  q_l^2(k)=\beta_l^2\,(k-\klp)\,(k-\klm); 
  ~~k_{l\pm}=\pm{mv\over\sqrt{\kappa}\pm v}\pm i\epsilon; 
\end{equation}
\begin{equation}
  q_t^2(k)=\beta_t^2\,(k-\ktp)\,(k-\ktm); 
  ~~k_{t\pm}=\pm{mv\over 1\pm v}\pm i\epsilon; 
\end{equation}
\begin{equation}
  k_{R\pm}=\pm{mv\over v_R \pm v}\pm i\epsilon.
\end{equation}
The quantities $\ktp,~\klp$, etc. locate the branch points of $\hat
G(k,m)$ in the complex $k$-plane.  Note that they are all proportional
to $mv$.  The quantities $k_{R\pm}$ are wave numbers for Doppler
shifted Rayleigh modes, and are also of order $mv$.

In the limit $k\gg mv$, the factors $\hat G^{(\pm)}$ look very similar
to the $\hat F^{(\pm)}$ given in (\ref{Ffactors}):
\begin{eqnarray}
  \hat G^{(+)}(k,m) &\approx & {\beta_l\,b(v)\over 
    \beta_t}\,(\epsilon+ik)^{1/2}, \cr 
  \hat G^{(-)}(k,m) & \approx  & (\epsilon-ik)^{1/2}.\label{Gapprox}
\end{eqnarray}
This will be a very useful approximation.

Given the $\hat G^{(\pm)}$ (not necessarily in the approximate forms
(\ref{Gapprox})), we can solve the Wiener-Hopf equation (\ref{WH1}))
for the shear stress and the shear displacement. The results are:
\begin{eqnarray}
  i(k-m)\hat\Sigma_S^{(-)}(k,m) & = &\cr
  -\hat G^{(-)}(k,m) \Biggl[{\hat E_m\over \hat G^{(-)}(m,m)}
  & - & i(k-m)\,\hat\Lambda_m^{(-)}(k,m)\Ym \cr 
  & + & i(k-m)\, \hat\Lambda_{cS}^{(-)}(k,m)\Biggr];
  \label{ikSigmaS} \\
  i(k-m) \hat U_S^{(+)}(k,m) & = &\cr
  {1\over \hat G^{(+)}(k,m)} \Biggl[{\hat E_m\over \hat G^{(-)}(m,m)}
  & + &  i(k-m)\,\hat\Lambda_m^{(+)}(k,m)\Ym \cr & - &   i(k-m)
    \,\hat\Lambda_{cS}^{(+)}(k,m) \Biggr].
  \label{ikUS}
\end{eqnarray}
Here,
\begin{equation}
  \hat\Lambda_m^{(\pm)}(k,m) =  \mp\int{dk'\over 2\pi i}\,
  {1\over k'-k\pm i\epsilon}\,\left[{\hat L(k',m)\,\hat 
      U_N^{(+)}(k'-m)\over \hat G^{(-)}(k',m)}\right],
\end{equation}
and
\begin{equation}
  \hat\Lambda_{cS}^{(\pm)}(k,m)=\mp\int{dk'\over 2\pi i}\,
  {1\over k'-k\pm 
    i\epsilon}\,{\hat\Sigma_{cS}^{(+)}(k',m)\over 
    \hat G^{(-)}(k',m)}.
\end{equation}

The next step is the analog of the derivation of the Barenblatt
condition in (\ref{BCondapp}). Just as we used the condition that the
zeroth-order normal stress be nonsingular at the tip to determine
$\Sigma_G$ in (\ref{SigmaG}) and $\ell$ in, for example,
(\ref{ellzeta}), so it appears now that we can choose $\Ym$ so that
the shear stress is nonsingular.  As we shall see, however, the
situation is not so simple as it seemed in CLN.

On the right-hand side of (\ref{ikSigmaS}), in the limit of large $k$,
the factor $\hat G^{(-)}(k,m)$ is proportional to $k^{1/2}$ and the
quantity in square brackets goes to a constant. Therefore, without
regularization, the shear stress would diverge like $|x|^{- 1/2}$.
Accordingly, we regularize the stress by requiring that the large-$k$
limit of the quantity in square brackets be zero, thus apparently
fixing the value of $\Ym$.  The result is: 
\begin{eqnarray}
  \hat E_m & = &\epm+im\Delep\Ym \cr
  &=&-\hat G^{(-)}(m,m)\,\int{dk\over 2\pi}\, {\hat L(k,m)\,\hat
    U_N^{(+)}(k-m)\over \hat G^{(-)}(k,m)}\,\Ym \cr && + \hat
  G^{(-)}(m,m)\,\int{dk\over 2\pi}\, {\hat\Sigma_{cS}^{(+)}(k,m) \over
    \hat G^{(-)}(k,m)}.
  \label{EmCond}
\end{eqnarray}

In principle, we can solve (\ref{EmCond}) for $\Ym$.  To do so, we
define the function $A(x)$ by 
\begin{equation}
  U_S(x)=U_N(x)\,\Theta(x)\equiv A(x)\,U_N(x)\,\Ym\,e^{imx}.
  \label{Adef}
\end{equation}
$A(x)$ needs to be defined only inside the cohesive zone, $0<x<\ell$.
It is also useful to define the function $u_S(x)$:
\begin{equation}
  U_S(x)\equiv u_S(x)\,\Ym\,e^{imx},~~~\hat 
  U_S^{(+)}(k,m)\equiv \hat u_S(k-m)\,\Ym.
\end{equation}
As argued in Section \ref{sec:summary}, the cohesive shear stress is
necessarily linear in $u_S(x)$ or, equivalently, in $A(x)$; thus we
can write [see (\ref{ScSeta})]
\begin{equation}
  \hat\Sigma_{cS}^{(+)}(k,m) = \Sigma_0\,\Ym \int_0^{\ell}\! dx
  \tilde\sigma_{cN}[U_N(x)]\tau_{cS}[A(x)]e^{-i(k-m)x},
\end{equation}
where, in the moving frame of reference,
\begin{equation}
  \tau_{cS}[A(x)]=(1+im\eta v)A(x)+\eta v\,{dA\over dx}.
\end{equation}

With this notation, (\ref{EmCond}) can be rewritten as the formally
exact expression for the response coefficient shown in (\ref{chi1}):
\begin{equation}
  -{\epm\over \Ym}=-\hat\chi_Y^{-1}(m,v) = im\Delep+\tilde{\cal
  D}_0(m,v) + \tilde{\cal D}_1(m,v)\label{chi1app} 
\end{equation}
where
\begin{equation}
  \tilde{\cal D}_0(m,v)=\int{dk\over 2\pi}\, {\hat G^{(-
      )}(m,m) \over \hat G^{(-)}(k,m)}\,\hat L(k,m)\,\hat 
  U_N^{(+)}(k-m);\label{D0}
\end{equation}
and
\begin{eqnarray}
  \tilde{\cal D}_1(m,v) &=& -\Sigma_0\,\int{dk\over 2\pi}\, {\hat
    G^{(-)}(m,m) \over \hat G^{(-)}(k,m)} \cr
  &\times& \!\! \int_0^{\ell} dx\,\tilde
  \sigma_{cS}[U_N(x)]\,\tau_{cS}[A(x)]\,e^{-i(k-m)x}. \label{D1}
\end{eqnarray}

It is useful to write $\tilde{\cal D}_0(m,v)$ in the form:
\begin{equation}
  \tilde{\cal D}_0(m,v)=\int_0^{\infty}\,dx\,h(x)\,U_N(x),\label{hU}
\end{equation}
where
\begin{equation}
  h(x)= \int {dk\over 2\pi}\,{\hat G^{(-)}(m,m)\over \hat 
    G^{(-)}(k,m)}\,\hat L(k,m)\,e^{-i(k-m)x}.
\end{equation}
The function $\hat L(k,m)$, defined in (\ref{Ldef}), is a homogeneous
function of $k$ and $m$ of order 2; that is, $\hat L(k,m)=m^2\,\hat
L(k/m,1)$.  Similarly, $\hat G^{(\pm)}(k,m)=\sqrt{m}\,\hat
G^{(\pm)}(k/m,1)$. Thus, $h(x)$ is a function only of $mx$.  For
$m\ell\ll 1$, $h(x)$ is very slowly varying on the scale $\ell$; that
is, it changes appreciably only for $x$ of order $1/m\gg\ell$.  As a
result, the overwhelmingly largest contribution to the integral in
(\ref{hU}) comes from the region outside the cohesive zone and, so
long as $mW$ remains large, we can use (\ref{Usqrtx}) to rewrite
(\ref{D0}) in the form 
\begin{equation}
  \tilde{\cal D}_0(m,v)\cong im(- imW)^{1/2}\Sigma_{N\infty}\,{\cal
    D}_0(v).
  \label{D0formapp}
\end{equation}
Here, ${\cal D}_0$ is a function only of $v$ [see CLN(6.13)]:
\begin{eqnarray}
  {\cal D}_0(v) &=& -{1 \over (-im)^{3/2}} \left[{2\over\kappa
      b(v)}\right]^{1/2}  \cr &\times& \int{dk \over 2\pi}{\hat
    G^{(-)}(m,m) \over \hat G^{(-)}(k,m)} 
  {\hat L(k,m) e^{-ik0}\over [\epsilon + i(k-m)]^{3/2}}. \label{D0exp}
\end{eqnarray}
(The convergence factor $e^{-ik0}$ is not strictly necessary here.)
Its limiting value as $v\to 0$ is given in (\ref{D00}). The easiest
way to derive (\ref{D0exp}) is, in effect, to step backwards and
replace $\sqrt x$ in (\ref{Usqrtx}) by
\begin{equation}
  \theta(x)\,\sqrt{x}=-{1\over 4\sqrt{\pi}} \int
  dk\,{e^{-ikx}\over(\epsilon-ik)^{3/2}}
\end{equation}
where $\theta(x)$ is the unit step function.

There is an important new point here.  Despite the appearance of
$U_N(x)$ in the formula for $\tilde{\cal D}_0(m,v)$, the latter
quantity is completely independent of what model we use for the
cohesive forces so long as we consider only perturbations whose
wavelengths $2\pi/m$ are much longer than $\ell$.

The formula (\ref{D1}) for $\tilde{\cal D}_1$ also can be simplified
for the case $m\ell\ll 1$. If we reverse the order of integration in
(\ref{D1}), so that it appears as a real-space integral in analogy to
(\ref{hU}), then the cohesive shear factor limits the integration to
the interior of the cohesive zone. As a result, the relevant values of
$k$ in the integrand in (\ref{D1}) are much larger than $mv$, and we
can use the approximation (\ref{Gapprox}) to write (\ref{D1}) in the
form:
\begin{eqnarray}
  \tilde{\cal D}_1(m,v) &\approx & - \hat G^{(-)}(m,m)\, \Sigma_0
  \cr & \times & \int_0^{\ell} \, {dx\over\sqrt{\pi\,x}}\,\tilde
  \sigma_{cS}[U_N(x)]\, \tau_{cS} [A(x)]\,e^{imx}.\label{D1approx}
\end{eqnarray}

At this point in our analysis, the remaining unknown ingredient of
$\hat\chi_Y(m,v)$ is the function $A(x)$ appearing in
(\ref{D1approx}).  According to (\ref{Adef}), $A(x)$ describes the
bending of the cohesive zone that is induced by the perturbing shear
stress. It is the central dynamical variable in this theory.

We must compute $A(x)$ by solving (\ref{ikUS}) for the
shear-displacement function $U_S(x)$. Equation (\ref{ikUS}) is an
inhomogeneous linear integral equation that determines $\hat
U_S^{(+)}(k,m)$, which appears explicitly on the left-hand side and
implicitly on the right-hand side {\it via} the factor
$\hat\Sigma_{cS}^{(+)}(k,m)$ in $\hat\Lambda_{cS}^{(+)}(k,m)$.  This
equation is best rewritten by using the regularization condition
(\ref{EmCond}) to eliminate the explicit $\hat E_m$, so that the
quantity in square brackets on the right-hand side automatically
decreases like $k^{-1}$ at large $k$. The result is [CLN(5.23)]:
\begin{eqnarray}
  i(k - m)\,\hat u_S^{(+)}(k - m) && =
  {1\over \hat G^{(+)}(k,m)} \!\int{dk'\over 2\pi} {k'- m\over k' - k
  + i\epsilon} \cr
  {1\over \hat G^{(-)}(k',m)} \Biggl[\Sigma_0 \!\int_0^{\ell} 
  \!\! dx \,&&\tilde\sigma_{cS}[U_N(x)] \tau_{cS}[A(x)] e^{-i(k'- m)x}
  \cr && -\hat L(k',m)\,\hat U_{N0}^{(+)}(k'-m) \Biggr].
  \label{CLN523}
\end{eqnarray}

It is useful --- indeed, essential --- to invert the Fourier transform
in (\ref{CLN523}) and study this equation in $x$- space. We find:
\begin{eqnarray}
  H(x) &=& {d\over dx}\,[A(x)\,U_N(x)] - \cr
  &&\Sigma_0\int_0^{\ell} \!\! dy\, \left[{\partial\over\partial y}
  K(x,y)\right] \tilde \sigma_{cS}[U_N(y)]\,\tau_{cS}[A(y)]
  \label{Aeqn}
\end{eqnarray}
Here,
\begin{equation}
  \label{Kdefapp}
  K(x,y)=\! \int{dk\over 2\pi i} {e^{i(k-m)x}\over \hat 
    G^{(+)}(k,m)}\! \int{dk'\over 2\pi} {1\over k-k'-
    i\epsilon} {e^{-i(k'-m)y}\over\hat G^{(-)}(k',m)},\label{K}
\end{equation}
and
\begin{equation}
  H(x)=\int_0^{\infty}dy\,h(x,y)\, {dU_N\over dy},\label{H}
\end{equation}
where
\begin{equation}
  \label{hdefapp}
  h(x,y)= \int{d\over 2\pi i} {e^{i(k-m)x}\over \hat 
    G^{(+)}(k,m)} \!\int{dk'\over 2\pi} {\hat L(k',m)\over k -
    k'-i\epsilon} {e^{-i(k'-m)y}\over \hat
    G^{(-)}(k',m)}. \label{hxy}
\end{equation}
In deriving (\ref{Aeqn}), we have integrated once by parts in the
second term on the left-hand side. This is legal because the upper
limit of integration, $\ell$, is still redundant;
$\tilde\sigma_{cN}[U_N(x)]$ is defined over the whole positive
$x$-axis and still may be assumed to vanish as smoothly as is
necessary for $x>\ell$. There is no difficulty at $x=0$ so long as the
stress remains finite there.

To make further progress we must develop approximate expressions for
$H(x)$ and $K(x,y)$ by exploiting the fact that the wavelength of the
perturbing stress is much larger than the length of the cohesive zone.
We begin by noticing that the expression (\ref{Kdefapp}) for $K(x,y)$
can be obtained from that for $h(x,y)$ in (\ref{hdefapp}) by replacing
$\hat L(k', m)$ by $1$.  Both expressions can be rewritten by defining
\begin{equation}
  \label{gammas}
  \Gamma^{(\pm)}(x,m) \equiv \int {dk \over 2\pi} {e^{\pm 
      ikx} \over \hat G^{(\pm)}(k,m)}.
\end{equation}
Note that $\Gamma^{(\pm)}(x,m) = 0$ for $x < 0$.  With the use of
these auxiliary functions we can now rewrite (\ref{hxy}) as
\begin{eqnarray}
  \label{hxy-gammas}
  h(x,y) & = & e^{-im(x - y)} \int_0^x dx' \, \Gamma^{(+)}(x') \cr
  &\times & \!\!\int dx'' \, L(x'',m) \Gamma^{(-)}(x' + x'' + y - x).
\end{eqnarray}
Here $L(x,m)$ is the inverse Fourier transform of $\hat L(k,m)$.  The
expression for $K(x,y)$ can be obtained from (\ref{hxy-gammas}) by
substituting a delta-function $\delta(x'')$ for $L(x'',m)$
\begin{equation}
  \label{Kxy-gammas}
  K(x,y) = e^{-im(x - y)} \int_0^x dx' \, \Gamma^{(+)}(x')
  \Gamma^{(-)}(x' + y - x).
\end{equation}
Approximate expressions for $\Gamma^{(\pm)}(x,m)$ can be derived for
$mx \ll 1$ by ignoring the small $k$ structure of $\hat
G^{(\pm)}(k,m)$.  Notable exceptions are the contributions to
$\Gamma^{(\pm)}$ from the poles at $k = k_{R\pm}$ which become large
as $v$ approaches $v_R$ even for small $m$.  We obtain
\begin{equation}
  \label{gamma(-)-approx}
  \Gamma^{(-)}(x,m) \cong {1 \over \sqrt{\pi x}} -{2i \over 
    \sqrt{\pi}} k_{R-}
  e^{ik_{R-} x} \int_0^{\sqrt{x}} dt\, e^{ik_{R-} t^2}.
\end{equation}
The analogous expression for $\Gamma^{(+)}(x,m)$ can be obtained
multiplying (\ref{gamma(-)-approx}) by $\beta_t/\beta_l b(v)$ and
replacing $k_{R-}$ by $-k_{R+}$.  In all the results reported here, we
have considered only velocities that are not close to the Rayleigh
speed that the second term in (\ref{gamma(-)-approx}) becomes
important.

With the assumption that $mv\ell\ll 1$, and the observation that both
$x$ and $y$ are small of order $\ell$ in (\ref{K}), we can use the
approximations (\ref{Gapprox}) to find 
\begin{equation}
  \label{Kapprox}
  K(x,y)\cong {\beta_t\over \pi b(v)\beta_l}\,\ln\left|{\sqrt 
      y +\sqrt x\over \sqrt y - \sqrt x}\right|\,e^{-im(x-y)}.
\end{equation}
For reasons discussed in Section \ref{sec:first}, we have left the factor
$e^{-im(x-y)}$ in (\ref{Kapprox}).  Remember that the approximations
(\ref{Gapprox}) are valid for small $mv$, that is, for small enough
$v$ at any value of $m\ell$.  The remaining $m$-dependence in
(\ref{Kapprox}) plays some role near the $v=0$ threshold.

As in the calculation of $\tilde{\cal D}_0$ in (\ref{D0form}), the
kernel $h(x,y)$ is a function only of $mx$ and $my$, and the integral
in (\ref{H}) is determined accurately, for $m\ell \ll 1$, by values of
$y$ outside the cohesive zone.  Therefore, in deriving an approximate
expression for $h(x,y)$, we may use the small $mx$ approximation only
for $\Gamma^{(+)}(x,m)$.  Returning to the Fourier transforms, we
obtain 
\begin{eqnarray}
  h(x,y) &\cong & {\beta_t\over b(v)\,\beta_l}\,e^{-im(x-y)} \int_0^x
  {dx'\over\sqrt{\pi x'}} \cr
  &\times & \! \int{dk\over 2\pi}\,{\hat L(k,m)\over
    \hat G^{(-)}(k,m)}\,e^{ik(x-x'- y)}.\label{hxyapprox}
\end{eqnarray}
Since the integral in (\ref{H}) is dominated by large values of $y$,
we use (\ref{Usqrtx}) to find 
\begin{equation} 
  H(x)\cong {3m^2\,(-im)^{1/2}\over 2\,\hat G^{(-)}(m,m)}\,
  {\Sigma_{N\infty}\sqrt{W}\over \sqrt\pi b(v)}\, {\cal
    D}_1(v)\,\sqrt{x} \label{Hform}
\end{equation}
where
\begin{eqnarray}
  {\cal D}_1(v) & = & {4\over 3m^2\,(-im)^{1/2}}\, 
  \left[{2\over\kappa\,b(v)}\right]^{1/2}\, 
  {\beta_t\over\beta_l} \cr
  & \times & \! \int{dk\over 2\pi}\,{\hat G^{(-)}(m,m) \over
    \hat G^{(-)}(k,m)}\, {\hat L(k,m)\,e^{-ik0}
    \over[\epsilon+i(k-m)]^{1/2}}.\label{D1exp}
\end{eqnarray}
In this case, the convergence factor is necessary. Before reversing
the order of integration over $k$ and $y$, which we have done here, we
must close the $k$-contour in the lower half plane in
(\ref{hxyapprox}) because we want $y > x- x'$ in the exponential
there.  Then, in (\ref{D1exp}), the factor $e^{-ik0}$ enforces this
closure rule. Unlike the situation in (\ref{D0exp}), we would get an
incorrect result here if we closed the contour in the upper half
plane. We have chosen the prefactors in (\ref{Hform}) and
(\ref{D1exp}) so that, as in the case of ${\cal D}_0(v)$, the function
${\cal D}_1(v)$ depends only on $v$ and has the same limit (\ref{D00})
as $v\to 0$.

Finally, we make the scaling transformations shown in (\ref{zetadef})
and (\ref{sigmadef}), and define $a(x)$ by 
\begin{equation}
  \label{Afactors} 
  A(x) = {m^2(-im\pi W \ell)^{1/2} \Sigma_{N\infty} \over 
    \hat G^{(-)}(m,m)\, \Sigma_0}\, {\cal D}_1(v)\, a(\xi). 
\end{equation}
The result is that (\ref{Aeqn}) becomes (\ref{aeqn}).

\section{Numerical methods}
\label{sec:numerics}

The most challenging numerical problems that we have faced in this
investigation occur in the inhomogeneous, linear, singular integral
equations of the form (\ref{aeqn}) or (\ref{aeqn3}), which are to be
solved for $a(\xi)$ or, equivalently, $\tau_{cS}[a(\xi)]$.  We also
have encountered nonlinear singular integral equations in the
steady-state problem described in Section \ref{sec:steady}.  Here,
however, the numerical problems are less severe; there are no weak
singularities that need to be resolved in order to select physically
admissible solutions.

Our basic strategy has been to replace the unknown function, say
$a(\xi)$, by a piecewise-constant approximant that takes on values
$a_i$ in intervals $[\xi_{i-1}, \xi_i]$, where $\xi_0=0$ and $\xi_N =
1$.  We have used regular as well as variable interval lengths.  Then
an integral equation of the form [compare (\ref{aeqn3})]
\begin{equation}
\label{inhomogeneqn}
  f(\xi)\,a(\xi)+\int_0^1 d\xi'\,Q(\xi,\xi')\,a(\xi') = g(\xi)
\end{equation}
can be evaluated at points $\xi= z_i \in [\xi_{i-1}, \xi_i]$ to become
a system of $N$ linear algebraic equations for the $N$ unknowns $a_i$.
We take the $z_i = (\xi_{i-1} + \xi_i)/2$ to be the centers of the
intervals. Thus,
\begin{equation}
  \label{eq:discretization}
  \sum_{j=1}^{N} C_{ij}\,a_j = B_i,
\end{equation}
where
\begin{equation}
  C_{ij} = f(z_i) \delta_{ij} + \int_{\xi_{j-1}}^{\xi_j}
  d\xi'\,Q(z_i,\, \xi'), \qquad B_i = g(z_i).
\end{equation}
In general, $C$ is a non-self-adjoint, complex matrix.

Because the matrix $C$ in all of our applications represents a
singular integral operator, it is ill-conditioned in the sense that at
least one of its eigenvalues vanishes in the continuum limit, $N \to
\infty$ .  The system of equations (\ref{eq:discretization}) has a
homogeneous solution in that limit; that is, (\ref{inhomogeneqn}) has
a solution with $g(\xi)=0$.  In general, particular solutions also
exist; therefore we can construct families of solutions by adding
arbitrary amounts of homogeneous solutions to particular solutions. Of
course, as described in Section \ref{sec:first-order-sol}, not all of
these solutions are physically acceptable.  In well-posed problems, we
expect to be able to exclude all but one of them.

For any large but finite $N$, $C$ may be a well-conditioned,
invertible matrix, but one of its eigenvalues is generally too small
for numerical purposes.  To deal with this difficulty, we perform a
singular-value decomposition \cite{svd} of $C_{ij}$ to write it in the
form
\begin{equation}
  \label{eq:SVD}
  C_{ij} = \sum_k U_{ik} W_k V^{\dag}_{jk},
\end{equation}
where the $W_k$ are the analogs of the eigenvalues of a
well-conditioned, i.e.~non-singular matrix.  (The term ``singular'',
as in ``singular matrix'' or ``singular-value decomposition'', refers
to non-invertibility of a matrix, i.e.~to the existence of null
eigenvectors, and not specifically to the singularity of an integral
operator in the sense of Muskhelishvili \cite{muskhelishvili}.)
The matrices $U$ and $V$ are unitary.  They are unique up to
permutations of columns (and corresponding elements of $W$) and their
columns are left and right eigenvectors of $C$ respectively.

We then use the singular value decomposition (\ref{eq:SVD}) to compute
the so-called ``minimum-norm'' solution:
\begin{equation}
  \label{eq:min_norm}
  a^{min}_i = \sum_{j,k}\,\!\!{}^{'} \, V_{ij} {1 \over W_j}
  U^{\dag}_{kj} B_k,
\end{equation}
where the symbol $\sum'$ means that, in the sum over $j$, we omit the
term for which $\lim_{N\to\infty} W_j = 0$.  In a finite matrix
problem, the vector $a^{min}$ is an exact particular solution of
(\ref{eq:discretization}) if the vector $B$ does not lie in the null
space of $C$, and this particular solution would have the smallest
possible norm, $\sum_i |a_i|^2$.  In our case, although $B$ generally
does have a component in the direction of the right null eigenvector
of $C$, the minimum-norm solution (\ref{eq:min_norm}) is guaranteed to
be well behaved for large $N$, and it does in fact converge to a
particular solution of (\ref{inhomogeneqn}).  We have checked this
convergence by computing the residual $\sum_i|\sum_j C_{ij} a^{min}_j
- B_i|^2.$ This residual was generally of the same order magnitude as
the smallest eigenvalue.  It vanished in the limit of large $N$ as
well.

Using this procedure for the various cases discussed in this paper, we
have been able to generate all members of what have always been
one-parameter families of solutions.  When both the homogeneous
solution $a^{null}_i$ and the minimum norm solution $a^{min}_i$ have
been found to be singular at $\xi=1$, the non-singular linear
combination has been obtained approximately, up to corrections of
order of the mesh interval squared, by the formula 
\begin{equation}
  \label{eq:linear_combination}
  a_i = a^{null}_i + {a^{null}_N - a^{null}_{N-1} \over a^{min}_N -
    a^{min}_{N-1}} a^{min}_i.
\end{equation}

We used a regular mesh with $N=400$ to compute the zeroes of $\Phi(m,
v)$ in the complex $m$-plane that are presented in Figures
\ref{fig:dugdale_poles}, \ref{fig:lambda_poles1},
\ref{fig:lambda_poles2} and \ref{fig:lambda_poles3}.  The detailed
study of the $\xi=1$ singularity in the null and minimum norm
solutions of the bending response integral equation \ref{aeqn3} was
done using a mesh with $N=800.$  The spacing of the mesh decreased as a
power law near $\xi=1$ to help achieve greater accuracy in resolving
the singularity.


\end{document}